\documentclass[traditabstract]{aa}
\usepackage{graphicx}
\usepackage[]{natbib}
\usepackage{enumerate}
\def\lunits{$\rm erg\,s^{-1}$~}
\def\funits{$\rm erg\,cm^{-2}\,s^{-1}$~}
\def\cunits{$\rm cm^{-2}~$}
\def\sax{{\it BeppoSAX~}}
\def\xmm{{\it XMM-Newton~}}
\def\chandra{{\it Chandra~}}

\def\micron{$\rm \mu m$~} 
\def\lxl6{$\rm L_X/L_{6\mu m}$~}

\begin{document}
\title{X-ray observations of highly obscured $\tau_{\rm 9.7\,\mu m}>1$ sources: An efficient method for selecting Compton-thick AGN ?}

%   \subtitle{ }

  \titlerunning{Highly obscured mid-IR sources as Compton-thick AGN}
    \authorrunning{I. Georgantopoulos et al.}

   \author{I. Georgantopoulos \inst{1,2},
           K. M. Dasyra \inst{3,4},
           E. Rovilos \inst{5},
           A. Pope \inst{6,7},
           Y. Wu \inst{8},
           M. Dickinson \inst{6},
           A. Comastri \inst{1},
           R. Gilli \inst{1},
           D. Elbaz \inst{3},
           L. Armus\inst{8} and 
           A. Akylas\inst{2}
                     }
   \offprints{I. Georgantopoulos, \email{ioannis.georgantopoulos@oabo.inaf.it}}

   \institute{INAF-Osservatorio Astronomico di Bologna, Via Ranzani 1, 40127, Italy
              \and 
              Institute of Astronomy \& Astrophysics,
              National Observatory of Athens, 
              Palaia Penteli, 15236, Athens, Greece
              \and
              Laboratoire AIM, CEA/DSM - CNRS - Universit\'{e} Paris Diderot,
              Irfu/Service d' Astrophysique, CEA Saclay, Orme des Merisiers,
              91191 Gif sur Yvette Cedex, France
              \and
              Observatoire de Paris, LERMA (CNRS:UMR8112), 
              61 Av. de l' Observatoire, F-75014, Paris, France
              \and
              Max Planck Institut f\"{u}r Extraterrestrische Physik,
              Giessenbachstra\ss e, 85748, Garching, Germany
              \and
              National Optical Astronomy Observatory,
              950 North Cherry Avenue, Tucson, AZ 85719, USA
              \and
              Department of Astronomy, University of Massachusets,
              Amherst, MA01003, USA
              \and 
              Spitzer Science Center, California Institute of Technology,
              MS 220-6, Pasadena, CA 91125, USA
                 }
   \date{Received ; accepted }

\abstract{Observations with the IRS spectrograph onboard Spitzer have found 
          many sources with very deep Si features at $\rm 9.7\,\mu m$, that have 
          optical depths of $\tau>1$. Since it is believed that a few of these
          systems in the local Universe are associated with Compton-thick active galactic nuclei 
          (hereafter AGN),
          we set out to investigate whether the presence of a strong Si
          absorption feature is a good indicator of a heavily
          obscured AGN. We compile X-ray spectroscopic observations available
          in the literature on the optically-thick ($\rm \tau_{9.7\,\mu m}>1$)
          sources from the $\rm 12\,\mu m$ IRAS Seyfert sample. We find that
          the majority of the high-$\tau$ optically confirmed Seyferts (six out of nine) in
          the $\rm 12\,\mu m$ sample are probably Compton-thick. Thus,
            we provide direct evidence of a connection between mid-IR
          optically-thick galaxies and Compton-thick AGN, with the success rate
          being close to 70\% in the local Universe. This is at least
          comparable to, if not better than, other rates obtained with photometric
          information in the mid to far-IR, or even mid-IR to X-rays.
          However, this technique cannot provide complete Compton-thick AGN
          samples, i.e., there are many Compton-thick AGN that do not display
          significant Si absorption, with the most notable example being
          NGC\,1068. After assessing the validity of the high $\rm 9.7\,\mu m$
          optical-depth technique in the local Universe, we attempt to
          construct a sample of candidate Compton-thick AGN at higher redshifts.
          We compile a sample of seven high-$\tau$ {\it Spitzer} sources in the 
          Great Observatories Origins Deep Survey (GOODS)
          and five in the {\it Spitzer} First-Look Survey. All these have been
          selected to have no  PAH features (EW$_{6.2\mu m}<$0.3\micron) to maximise the
          probability that they are bona-fide AGN. Six out of the seven GOODS sources
          have been detected in X-rays, while for the five FLS sources only X-ray flux upper limits are available.
           The high X-ray luminosities ($\rm L_X>10^{42}$ \lunits) of the detected GOODS sources 
           corroborates that these are AGN. For FLS, ancillary optical spectroscopy reveals hidden nuclei in
         two more sources. SED fitting can support the presence of an AGN in the vast majority of  sources.
          Owing to the limited photon statistics, we cannot derive 
           useful constraints from X-ray spectroscopy on whether these sources are 
            Compton-thick.   However, the low $\rm L_X/L_{6\,\mu m}$
          luminosity ratios, suggest that at least four out of the six detected sources in GOODS 
           may be associated with Compton-thick AGN. 
        }
   \maketitle
%
%________________________________________________________________

\section{Introduction} 

Hard X-rays (2-10\,keV) are extremely efficient in detecting AGN, as they can
penetrate large columns of dust and gas. This allowed the {\it Chandra} and
{\it XMM-Newton} missions to map the AGN universe with unprecedented detail. In
particular, about 90\% of the X-ray background has been resolved
\citep{Alexander2003,Luo2008}, revealing a sky density in the CDF-N of
$\rm >5000\,deg^2$ \citep{Bauer2004}. The majority of these sources are
obscured AGN, presenting column densities $>10^{22}$\,\cunits
\citep[e.g.][]{Tozzi2006,Akylas2006}.

However, even the hard X-ray surveys may be missing a substantial fraction of
the most heavily obscured sources, the Compton-thick AGN, which have column
densities $\rm >10^{24}\,cm^{-2}$. Although this population remains elusive
\citep[see][for a review]{Comastri2004}, there is concrete evidence for its
presence. The peak of the X-ray background at 20-30\,keV
\citep[e.g.][]{Frontera2007,Churazov2007,Moretti2009} can be reproduced only by
invoking a significant number of Compton-thick sources at moderate redshifts.
 However, the exact density of Compton-thick sources 
required by X-ray background synthesis models still remains an open issue
\citep*{Gilli2007,Sazonov2008,Treister2009}.
Additional evidence of a numerous Compton-thick population
comes from the directly measured space density of black holes in the local
Universe \citep[see][]{Soltan1982}. It is found that the black hole space
density is a factor of 1.5-2 higher than that predicted from the X-ray
luminosity function \citep{Marconi2004,Merloni2008}. The exact number
depends on the assumed efficiency of the conversion of gravitational energy to
radiation. Direct searches in the ultra-hard 20-70\,keV band by {\it Swift} and
{\it INTEGRAL} did not detect large numbers of Compton-thick sources
\citep[e.g.][]{Ajello2008,Tueller2008,Paltani2008,Winter2009}. Nevertheless, it
is possible that even these ultra-hard surveys are biased against the most
heavily obscured reflection-dominated Compton-thick sources \citep{Burlon2010}.
Owing to the limited imaging capabilities of these missions, the flux limit
probed is very bright ($\sim 10^{-11}$\,\funits) allowing only the detection of
AGN in the local universe. At higher redshifts, some  Compton-thick
 AGN have been reported in the deepest \xmm and \chandra observations 
  in the \chandra deep fields \citep[]{Tozzi2006,Georgantopoulos2009,Comastri2011,Feruglio2011}. 

 Mid-IR wavelength observations have attracted much attention because they provide an
alternative way of detecting heavily obscured systems. This is because the
absorbed radiation by circumnuclear dust is re-emitted in the IR part of the
spectrum. \citet{MartinezSansigre2005} argue that a population of bright
$\rm 24\,\mu m$ AGN with no $\rm 3.6\,\mu m$ detections is as numerous as
unobscured QSOs at high redshift ($z>2$). On the basis of X-ray stacking analysis, 
\citet{Daddi2007, Fiore2008, Georgantopoulos2008, Treister2009b, Eckart2010, Donley2010} 
 propose that a fraction of infrared excess, $\rm 24\,\mu m$-bright
sources are associated with Compton-thick AGN. These sources are found at
high redshift ($z\sim2$), and their contribution to the X-ray background is
expected to be small \citep[$<$1\%;][]{Treister2009}. In contrast, the bulk of the
contribution to the X-ray background is produced at redshifts $z\sim0.7-1$
\citep{Gilli2007}.

At high redshifts, mid-IR spectroscopy with {\it Spitzer}-IRS, has detected a
number of sources with large columns of obscuring material, as inferred from
their $\rm 9.7\,\mu m$ Si features \citep[$\tau >1$; hereafter called
high-$\tau$ sources;][]{Dasyra2009}.  These systems could be associated with 
 Compton-thick AGN. A local analog system is the nearby ULIRG
NGC\,6240 \citep{Armus2006}, which is well known in X-ray wavelengths to host a
Compton-thick AGN \citep{Vignati1999}. There is little information available at
X-ray wavelengths for these high-$\tau$ systems at higher redshift
\citep{Bauer2010}.
 
The primary goal of this paper is to investigate whether an efficient way to
identify Compton-thick AGN is indeed to look for sources with deep silicate absorption
at $\rm 9.7\,\mu m$. The structure of the paper is as follows: we first compile a local
sample of high optical-depth ($\rm \tau_{9.7\,\mu m}>1$) sources from the
\citet{Wu2009} flux-limited {\it Spitzer}-IRS observations of the
$\rm 12\,\mu m$ Seyfert sample of \citet{Rush1993}.  We investigate the
X-ray spectra available in the literature, to deduce how many of these
high-$\tau$ objects are heavily obscured or Compton-thick. 
Finally, we compile a sample of high-$\tau$ AGN at higher redshift with
available X-ray observations, using the {\it Chandra} Deep Fields, as well as
the {\it Spitzer} FLS sample. Although these sources are faint in X-ray
wavelengths (and thus it is difficult to derive with great certainty their X-ray spectral
properties), we attempt to estimate whether they are heavily absorbed from
their $\rm L_x/L_{6\,\mu m}$ luminosity ratio.

We adopt $\rm H_o=70\,km\,s^{-1}\,Mpc^{-1}$, $\rm\Omega_{M}=0.3$, and 
$\Omega_\Lambda=0.7$ throughout the paper.

\section{Sample selection}

\subsection{The local sample}

 We employ a sample of local galaxies with a deep Si
absorption feature in the mid-IR, that are known to definitely host an AGN and
also have X-ray data, so that the fraction of Compton-thick AGN among them can
be quantified. An ideal data set for this purpose is the {\it Spitzer} IRS spectroscopic
sample of \citet{Wu2009}. These authors present low-resolution {\it Spitzer}
$\rm 5.5-35\,\mu m$ spectra for 103 galaxies from the $\rm 12\,\mu m$ Seyfert
sample \citep{Rush1993}. This is a complete, unbiased, flux-limited sample of
local Seyfert galaxies, selected from the {\it IRAS} Faint Source Catalog by
means of optical spectroscopy. 

To identify sources with high optical depth at $\rm 9.7\,\mu m$
($\tau_{\rm 9.7}$) due to Si absorption, we use the 6 and $\rm 13\,\mu m$
continuum flux to interpolate the intrinsic AGN unobscured flux at
$\rm 9.7\,\mu m$ in a similar way to \citet{Spoon2007}. We subsequently measure
$\tau_{9.7}$ by calculating the natural logarithm of its observed to intrinsic
value. For screen extinction, this corresponds to $-\tau_{9.7}$. For sources
without spectral coverage at $\rm 13\,\mu m$, we use the 6 and $\rm 7\,\mu m$
continuum values to extrapolate at $\rm 9.7\,\mu m$. We reject all sources with
optical depths smaller than one. Nonetheless, various spectral fitting techniques can
lead to different $\tau_{\rm 9.7}$ values due to different continuum
assumptions, see e.g. the small differences in our measurements of $\tau_{9.7}$
and those derived by \citet{Bauer2010} in the FLS sources. For this reason, we
also run {\sl PAHFIT} \citep[see][]{Smith2007} for all sources for which it is
possible, using the parameters presented in \citet{Dasyra2009}. We confirm that
all of the sources selected with our primary technique also have
$\tau_{\rm 9.7}>1$ based on {\sl PAHFIT}. The above re-analysis of the
\citet{Wu2009} {\it Spitzer} spectra shows that $\tau_{9.7}>1$ for eleven AGN 
(see Table\,\ref{irlocal}). The IRS spectra of many of these
sources were also discussed in \citet{Armus2007}.

\begin{table*}
\centering
\caption{{\it Spitzer} IRS properties of the high-$\tau$ $\rm 9.7\,\mu m$ AGN in the $\rm 12\,\mu m$ sample}
\label{irlocal}
\begin{tabular}{ccccccc}
\hline\hline
Name                 &  z    & Type & $\tau$  & $\rm log[\nu L_{6\,\mu m}]$ &  AGN& EW(6.2)  \\
 (1)                 &  (2)  & (3)  &   (4)  &   (5)     & (6) & (7)                           \\
\hline
Mrk\,938               & 0.020 & Sy2  & 1.2                 & 10.09    & 0 & 0.440                \\
NGC\,1125            & 0.011 & Sy2  & 1.0               &  9.12 & 0.81 & 0.258                  \\
I\,08572+3915    & 0.058 & Sy2  & 3.5              & 11.51    & 0.98 & $<$0.021        \\
UGC\,5101            & 0.039 & Sy1  & 1.4             & 10.52    & 0 & 0.229          \\
NGC\,3079            & 0.004 & Sy2  & 1.3             &  9.10      & 0 & 0.458        \\
Mrk\,266             & 0.028 & Sy2  & 1.0            & 10.03   &0 & 0.608      \\
Mrk\,273             & 0.038 & Sy2  & 1.7             & 10.51  &0 & 0.192        \\
Arp\,220             & 0.018 & Sy2  & 2.4             &  9.86  &0 & 0.344       \\
I\,19254-7245     & 0.062 & Sy2  & 1.2         & 11.03  &0.88 & 0.064        \\
NGC\,7172            & 0.009 & Sy2  & 1.9     &  9.75  &0.2 & 0.045        \\
NGC\,7582            & 0.005 & Sy2  & 1.0    &  9.56  & 0 & 0.274         \\ 
 \hline\hline
\end{tabular}
\begin{list}{}{}
\item The columns are: (1) Name;
                       (2) Redshift;
                       (3) Optical AGN type;
                       (4) Optical depth at $\rm 9.7\,\mu m$; 
                         (5) Logarithm of $\nu L_\nu$ IR monochromatic luminosity;
                           at $\rm 6\,\mu m$ in units of solar luminosity;
                           (6) Fraction of AGN contribution  at 6$\rm \mu m$ according to the spectral decomposition based on 
                            broad-band mid-IR photometry available from NED (see section 3.2.3 for details of the models used); 
                           (7) Equivalent-width of PAH 6.2 $\rm \mu m$ feature in units of $\rm \mu m$ taken 
                            from \citet{Wu2009}. 
\end{list}
\end{table*}

\begin{table*}
\centering
\caption{X-ray properties of the high-$\tau$ $\rm 9.7\,\mu m$ AGN in the $\rm 12\,\mu m$ sample}
\label{xlocal}
\begin{tabular}{ccccccc}
\hline\hline
Name                 &  z    &  $\rm N_H$ & $ \rm log[L_x]$  & X-ray Ref.         & Mission                 & Comment \\
 (1)                 &  (2)  & (3)  &   (4)  &   (5)     & (6)             & (7)                 \\
\hline
Mrk\,938               & 0.020 &   $40$    &  42.30          & 1,16                & {\tiny \it XMM}               & -       \\
NGC\,1125            & 0.011 &    -      &  41.97          & 2                & {\tiny \it SWIFT}             & -       \\
I\,08572+3915    & 0.058    &    -      &  41.30          & 3               & {\tiny \it Chandra}           & -       \\
UGC\,5101            & 0.039 &   140     &  41.67         & {\tiny 4,6,7,8,11,16} & {\tiny \it Chandra}/{\tiny \it XMM} & a       \\
NGC\,3079            & 0.004 &   200     &  40.25         & 5,16                 & {\tiny \it XMM}               & a       \\
Mrk\,266  & 0.028 & $>$160    &  41.7               & 1,6          & {\tiny \it XMM}/{\tiny \it Chandra} & a     \\
Mrk\,273             & 0.038   &    40     &  42.40        & 10,16                & {\tiny \it Chandra}           & -       \\
Arp\,220             & 0.018  & $>100$    &  40.96        & 7,12,14               & {\tiny \it XMM}               & b,c     \\
I\,19254-7245     & 0.062   & $>100$    &  42.57       & {\tiny 7,8,13}    & {\tiny \it BeppoSAX}          & a       \\
NGC\,7172            & 0.009   &     8     &  42.20          & 9                & {\tiny \it BeppoSAX}          & -       \\
NGC\,7582            & 0.005   &   160     &  42.00        & 9,15,17           & {\tiny \it BeppoSAX}          & a       \\ 
 \hline\hline
\end{tabular}
\begin{list}{}{}
\item The columns are: (1) Name.
                       (2) Redshift.
                        (3) X-ray column density in units of $10^{22}$\,\cunits.
                       (4) Logarithm of the obscured X-ray luminosity in the
                           2-10\,keV band in units of \lunits.
                         (5) X-ray spectroscopy reference:
                           1) \citet{Guainazzi2005}
                          2) \citet{Cusumano2010}
                         3) \citet{Iwasawa2009}
                          4) \citet{Imanishi2003}
                           5) \citet{Akylas2009}
                          6) \citet{Brassington2007}
                         7) \citet{Iwasawa2005}
                        8) \citet{Braito2003}
                          9) \citet{Dadina2008}
                           10) \citet{Iwasawa2011}
                          11) \citet{Gonzalez2009}
                         12) \citet{Ptak2003}
                        13) \citet{Georgantopoulos2010}
                         14) \citet{Clements2002}
                          15) \citet{Bianchi2009}
                          16) \citet{Brightman2010}.
                       (6) X-ray mission.
                      (7) X-ray criterion on which the source is classified as
                           Compton-thick:
                           a. detection of the absorption turnover
                           b. high-equivalent width Fe$K\alpha$ line
                           c. Flat spectrum (see text for details). 
\end{list}
\end{table*}

\subsection{High-$\tau$ $\rm 9.7\,\mu m$ AGN at higher redshift}

To identify sources in the distant Universe that are deeply obscured in the
mid-IR, we used all the {\it Spitzer}-IRS spectra available for the 4\,deg$^2$
field of the First Look Survey (FLS), as well as for the CDF-N and CDF-S
($\sim$900\,arcmin$^2$ in total) of the Great Observatories Origins Deep
Survey (GOODS). The two surveys are complementary in targeting sources for our
analysis. The FLS is a shallow $\rm 24\,\mu m$ survey, whose spectra were
limited to a depth of 0.9\,mJy \citep{Yan2007,Sajina2007,Dasyra2009}, targeting
luminous high-z galaxies. This is ideal for identifying many
high-$\tau$ systems, as the highest amounts of obscuration are found in the
sources with the highest infrared luminosity \citep*{Imanishi2010}. 
 However, the GOODS area has the most
 sensitive mid-IR and X-ray observations available.  
We select again sources with $\tau_{9.7\mu m}>1 $, using the methodology described above. 

\begin{figure*}
\rotatebox{0}{\includegraphics[width=5cm]{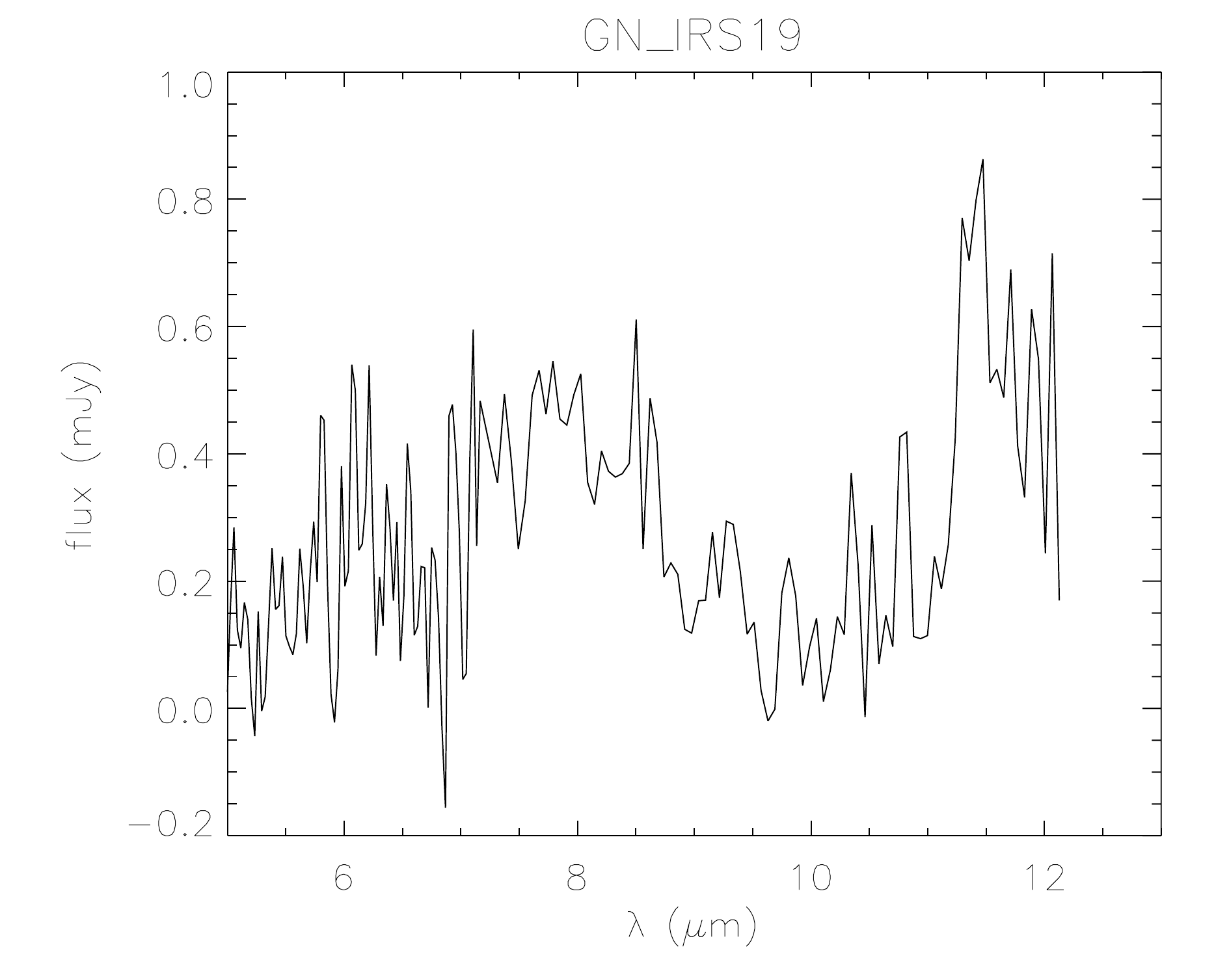}}
\rotatebox{0}{\includegraphics[width=5cm]{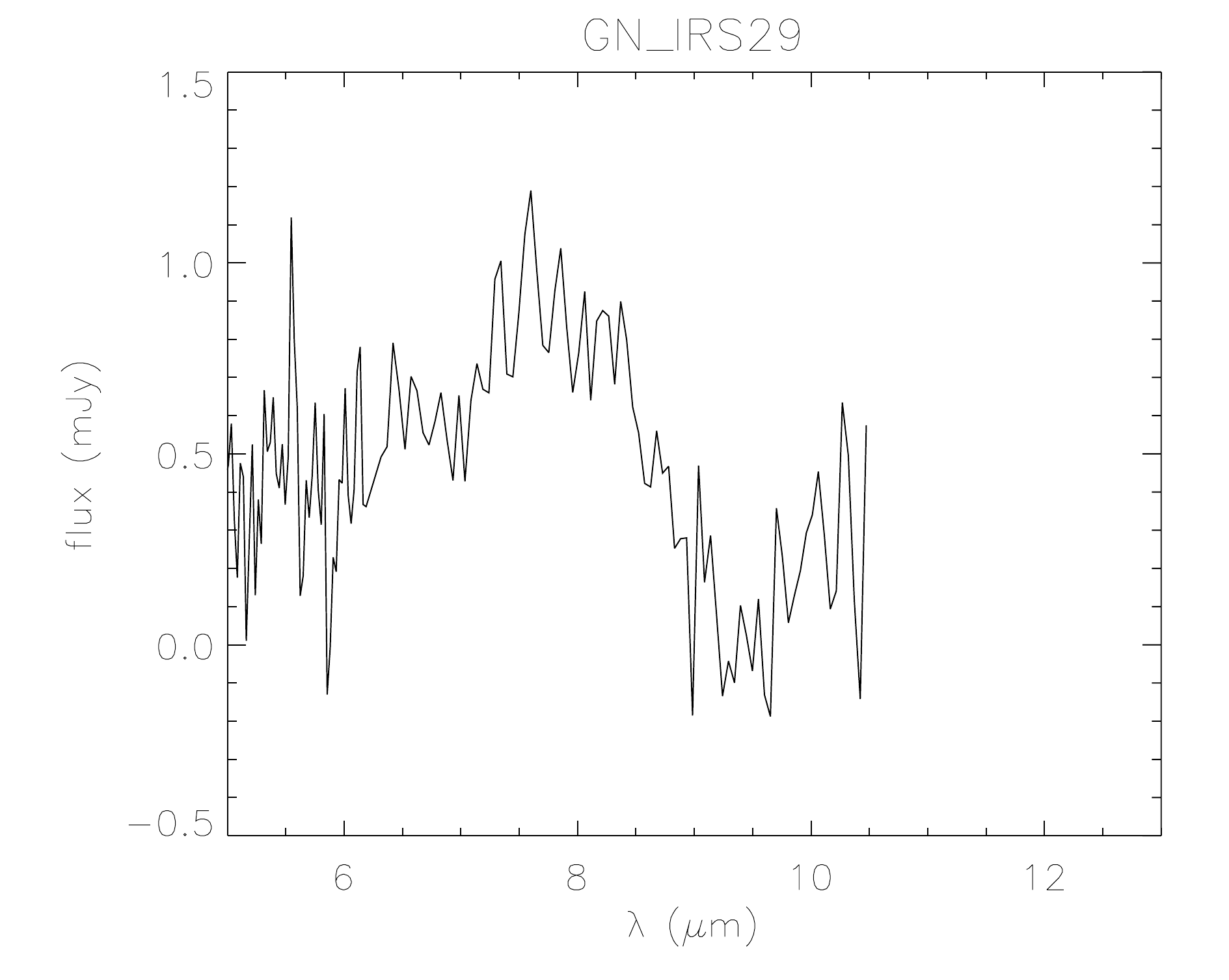}}
\rotatebox{0}{\includegraphics[width=5cm]{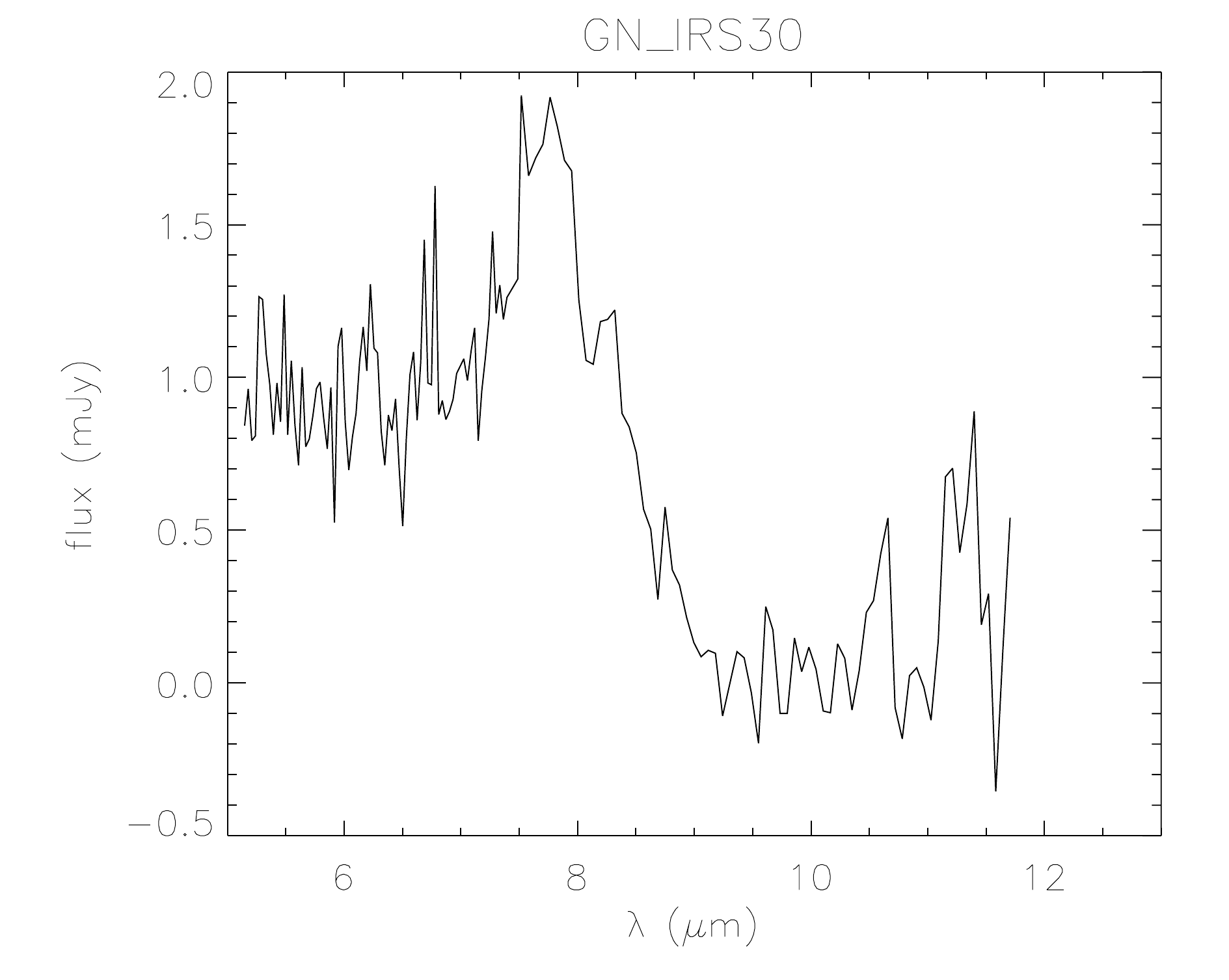}}\hfill \\
\rotatebox{0}{\includegraphics[width=5cm]{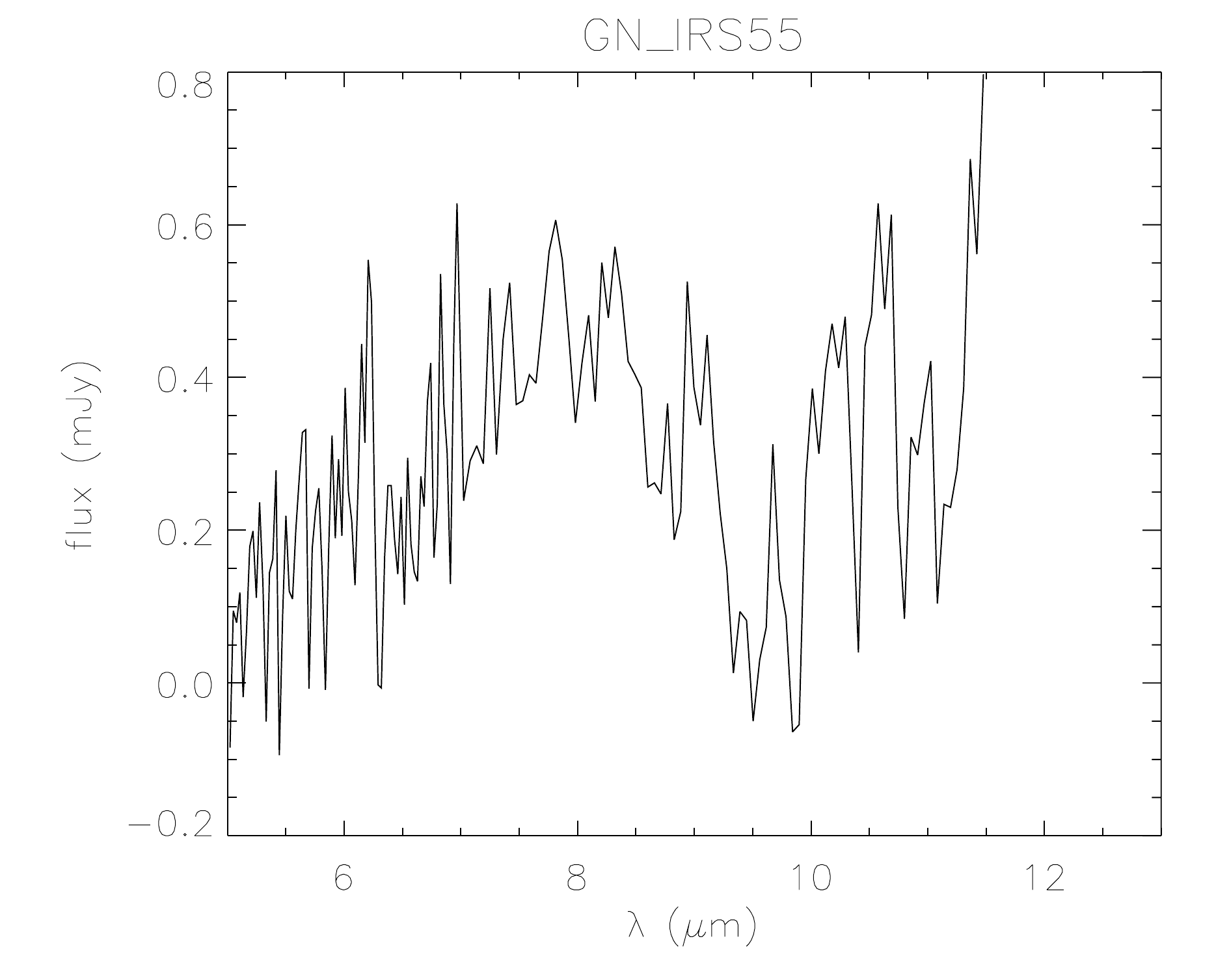}}
\rotatebox{0}{\includegraphics[width=5cm]{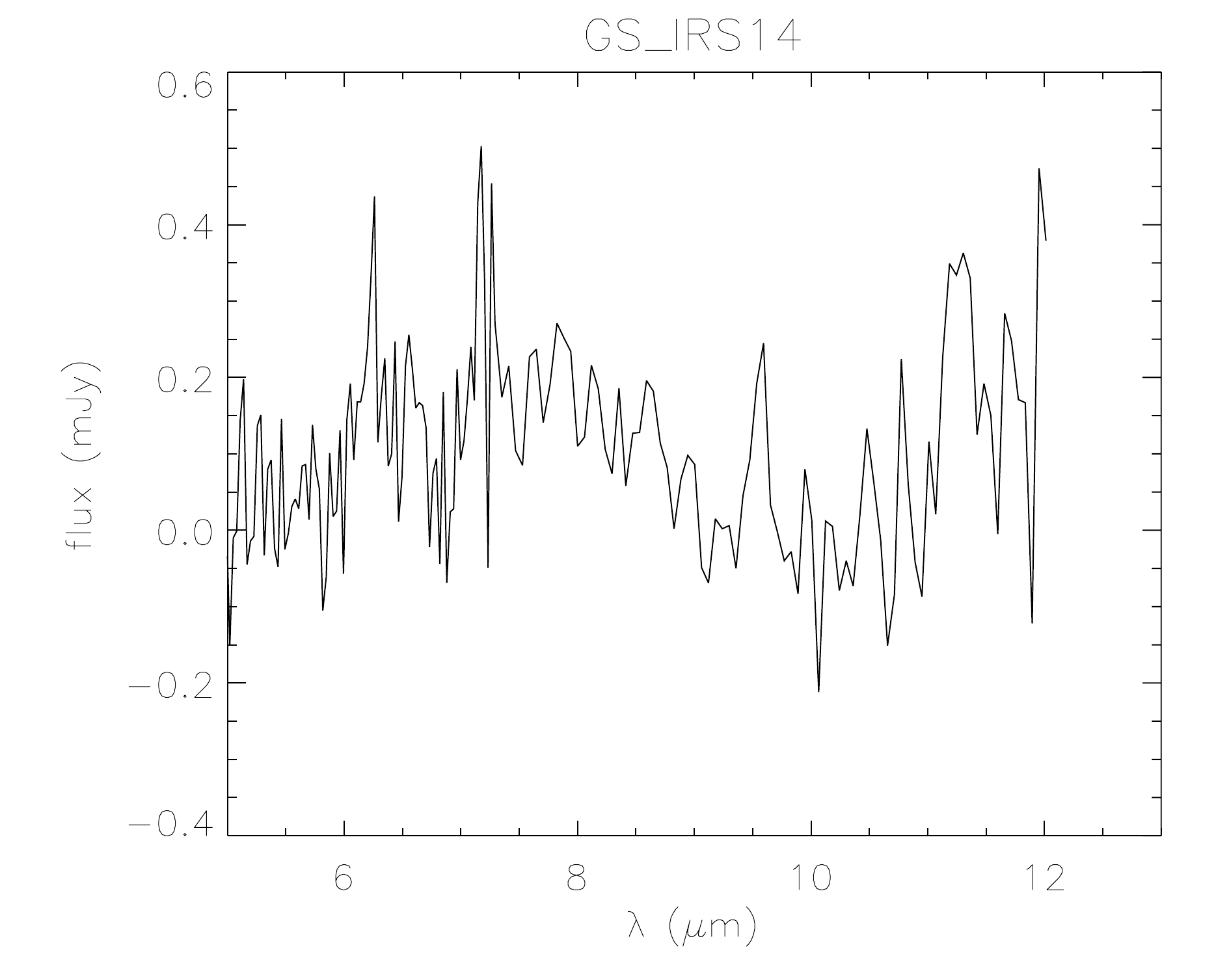}}
\rotatebox{0}{\includegraphics[width=5cm]{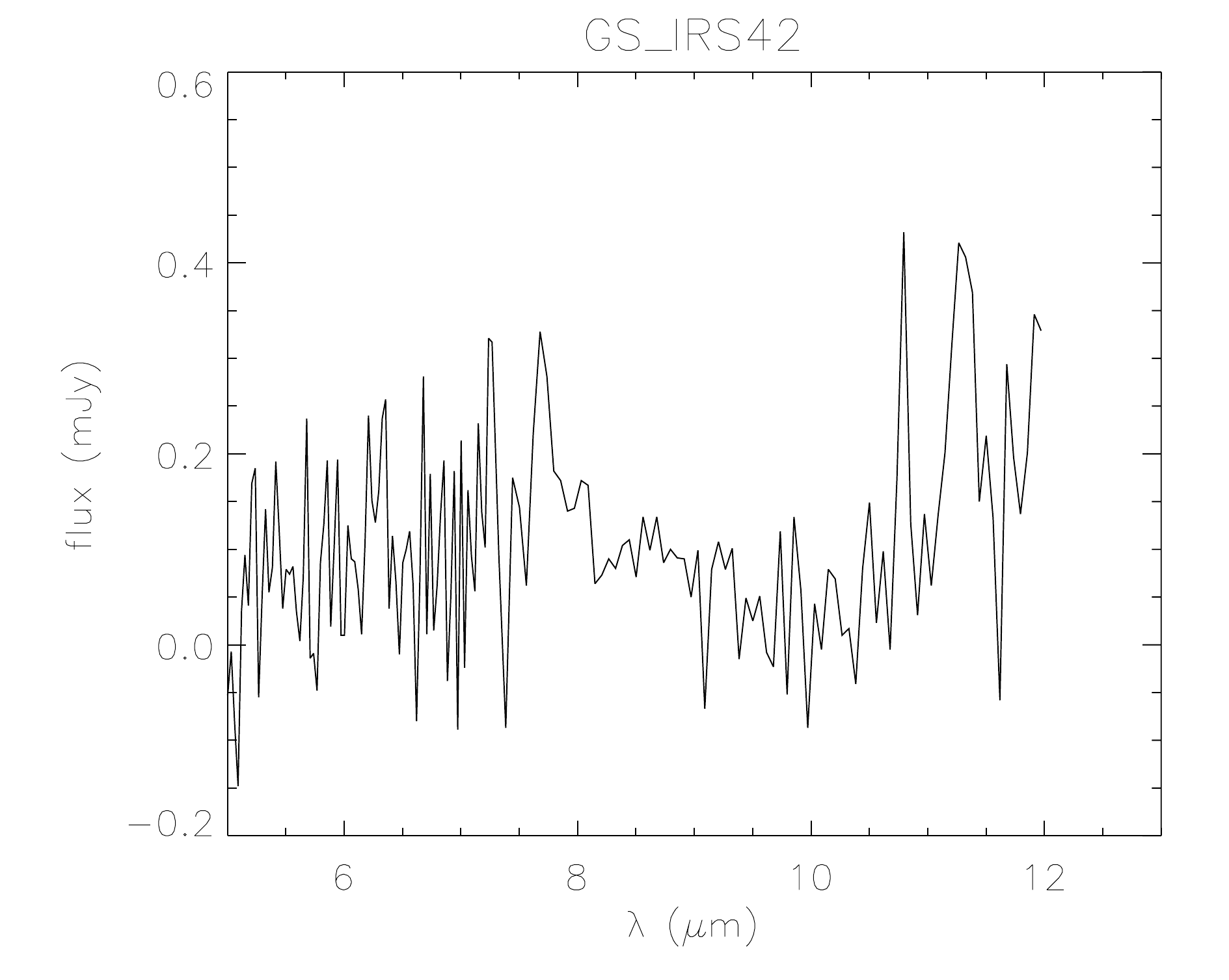}}\hfill \\
\rotatebox{0}{\includegraphics[width=5cm]{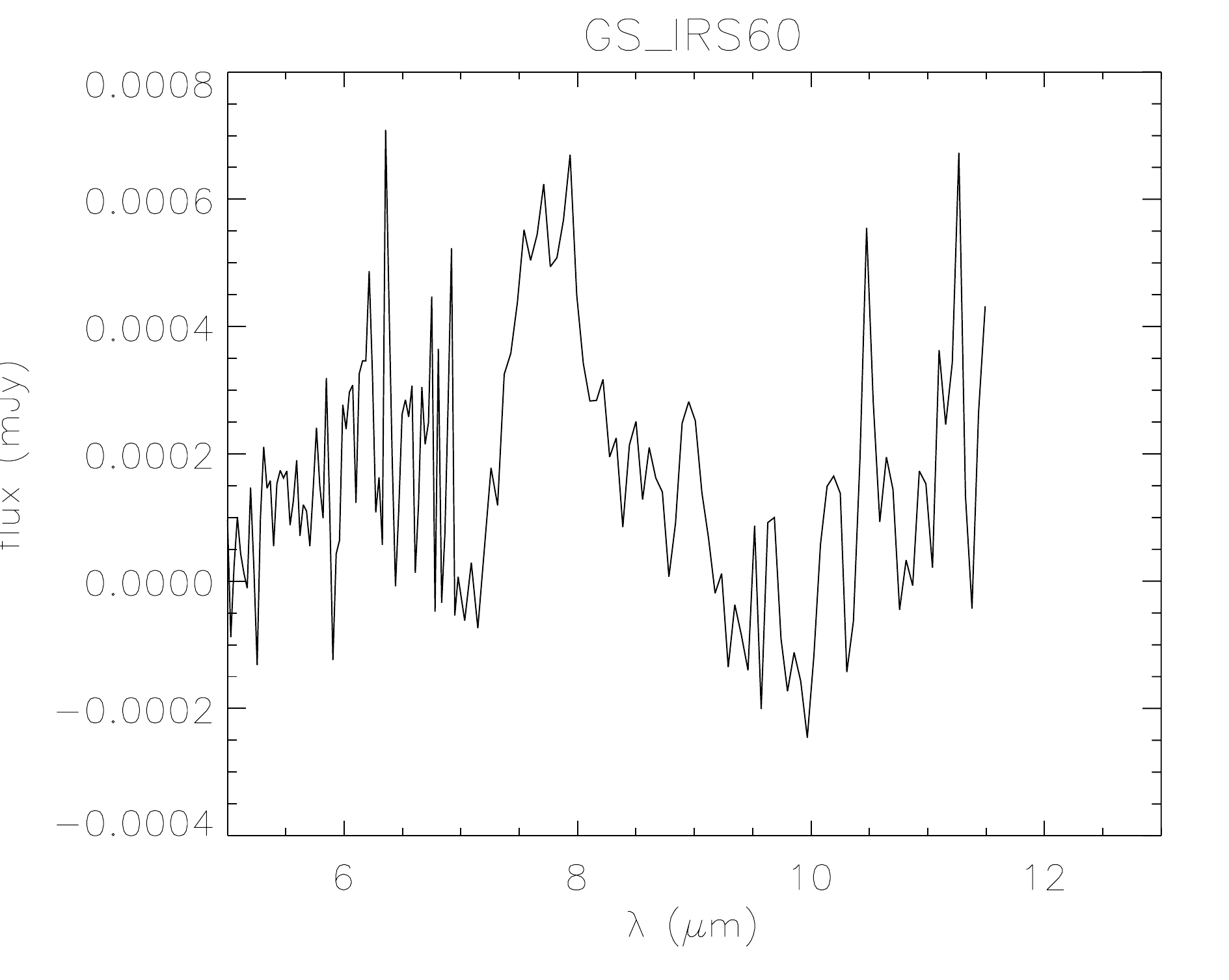}}
\rotatebox{0}{\includegraphics[width=5cm]{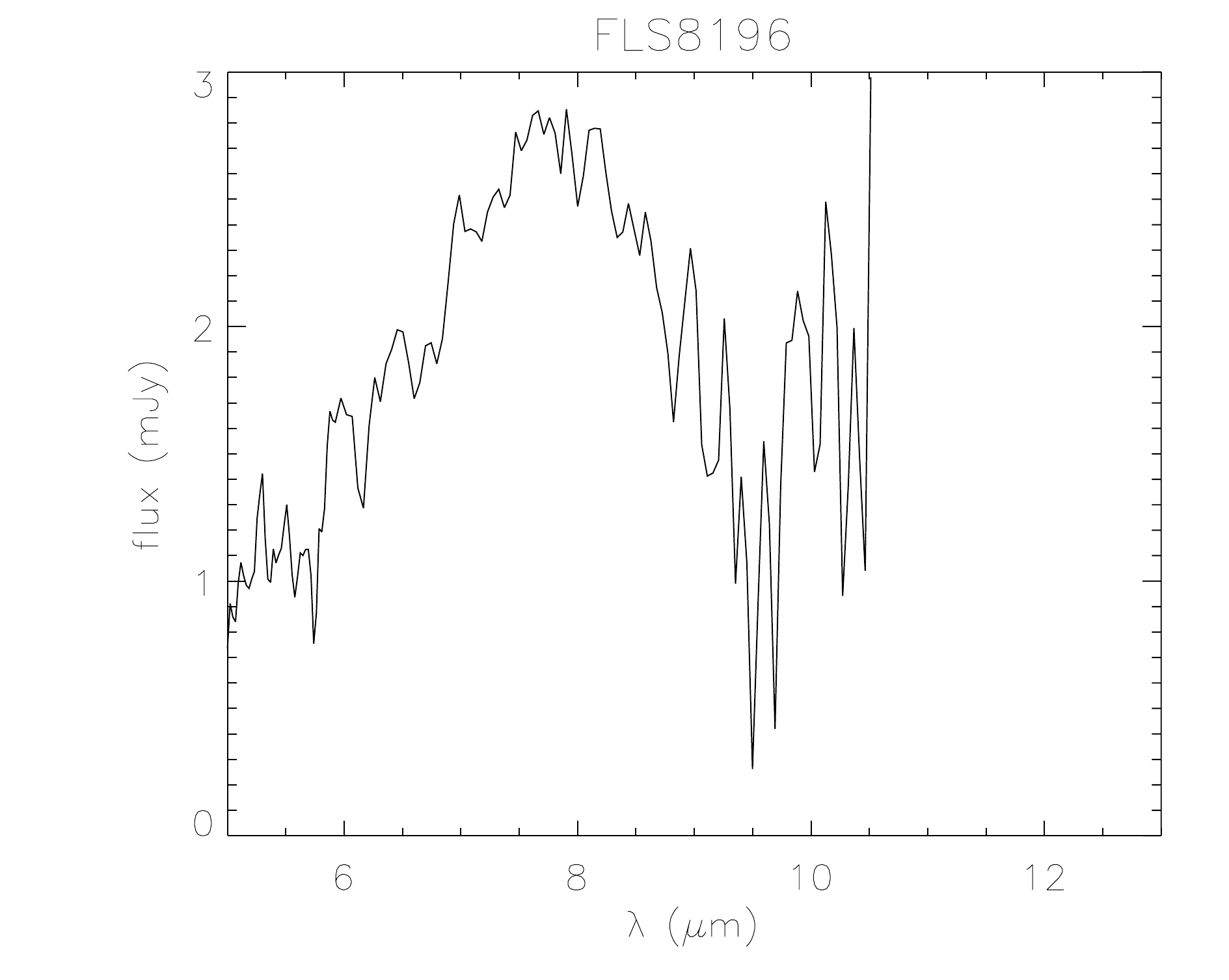}}
\rotatebox{0}{\includegraphics[width=5cm]{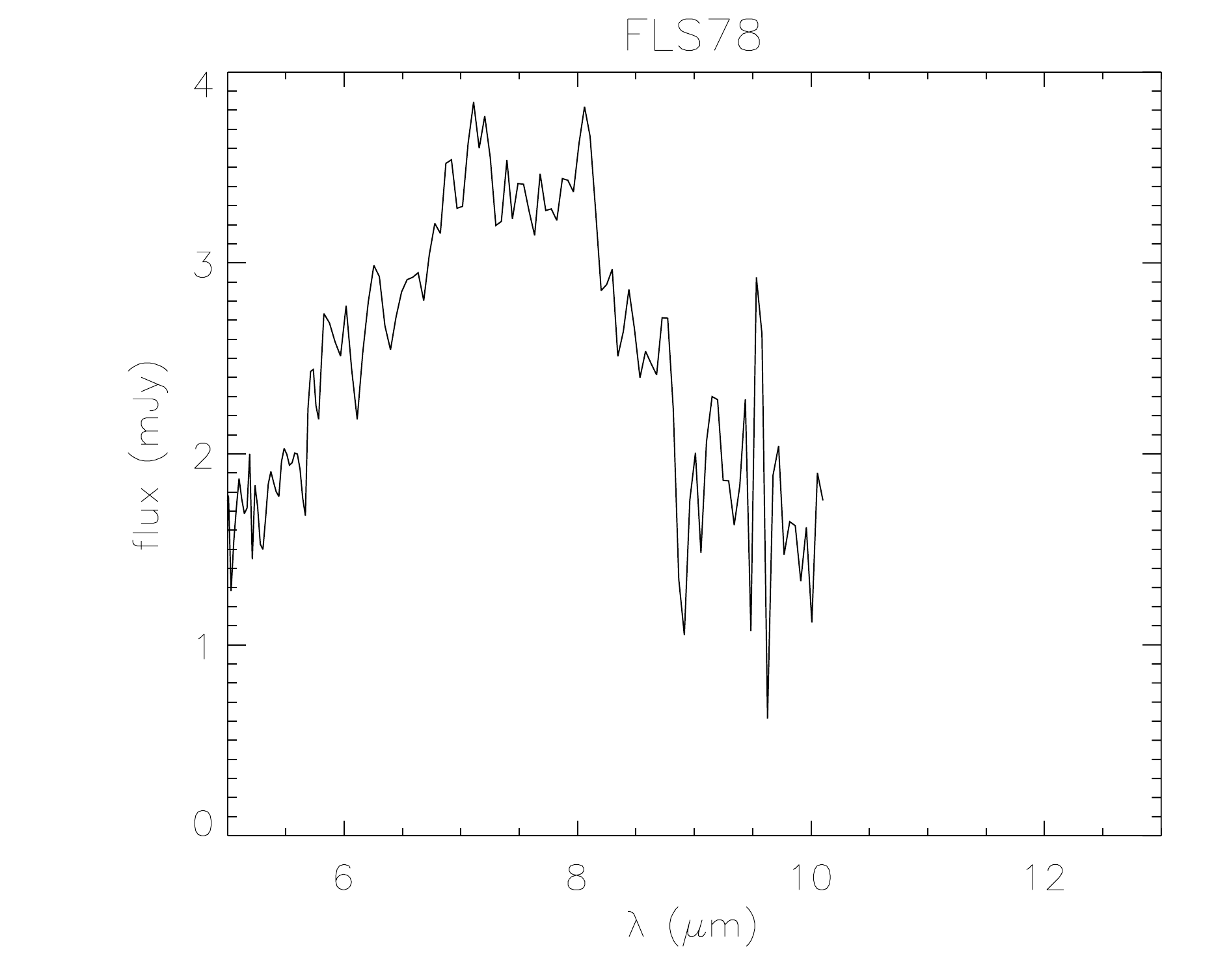}}\hfill \\
\rotatebox{0}{\includegraphics[width=5cm]{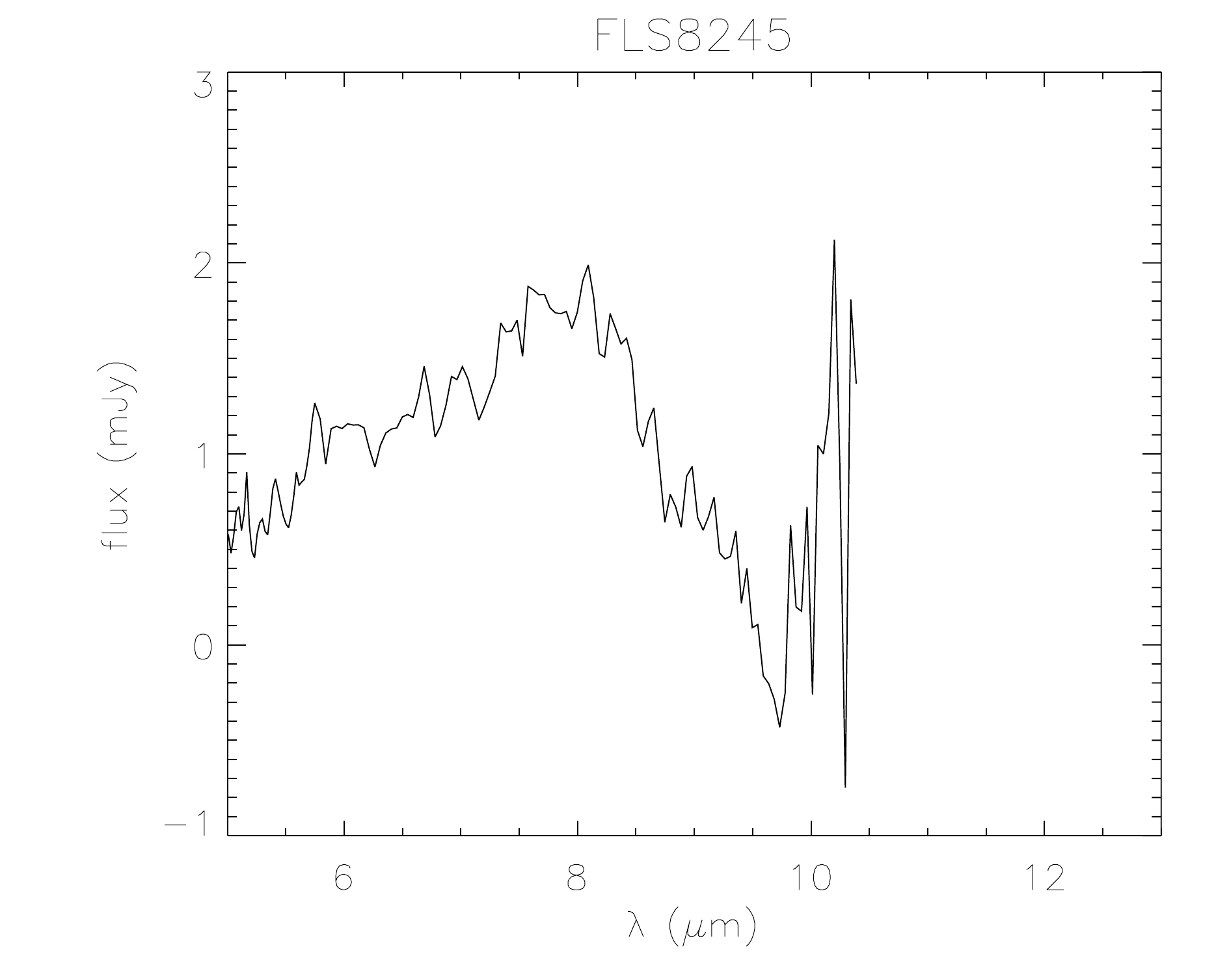}}  
\rotatebox{0}{\includegraphics[width=5cm]{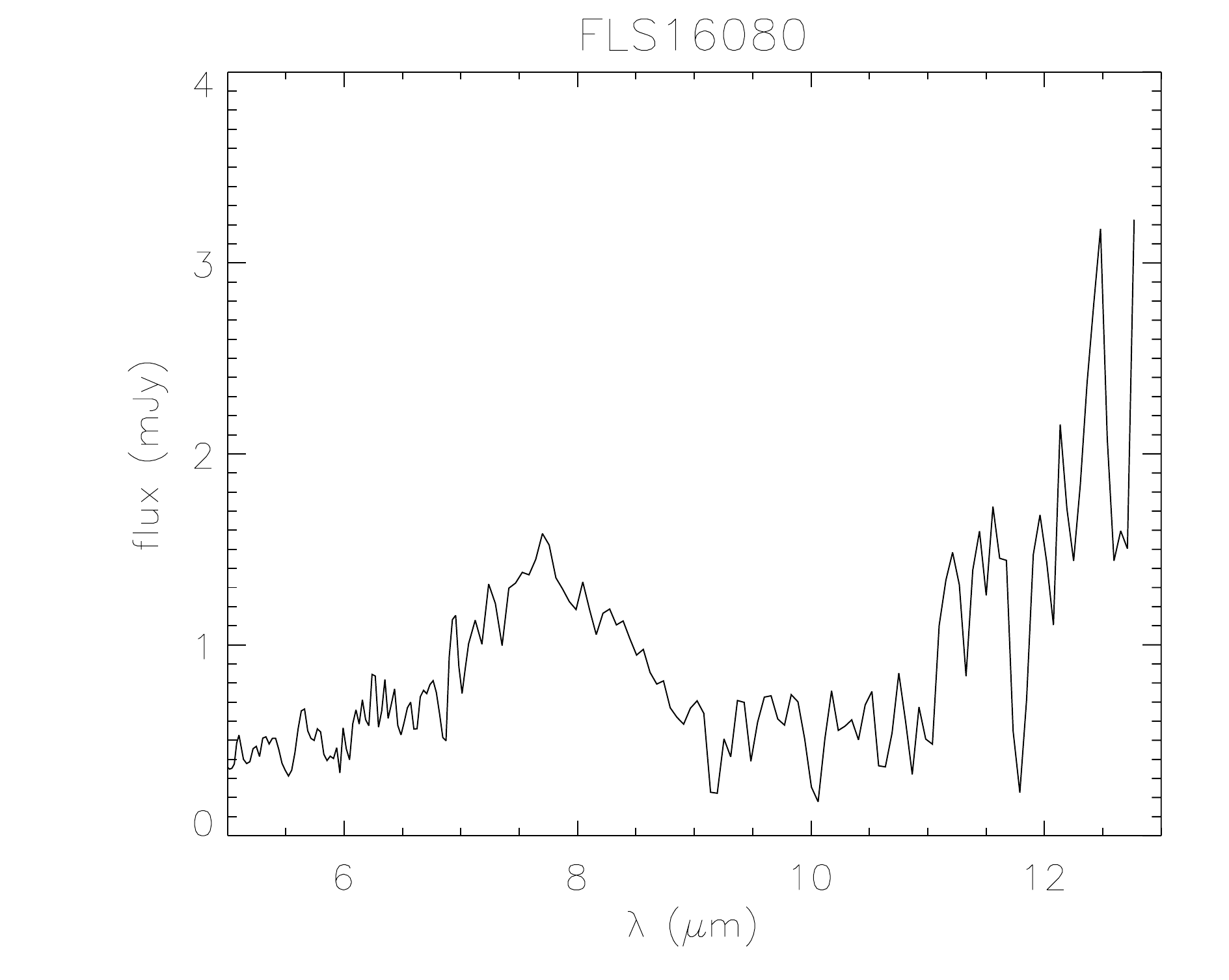}}
\rotatebox{0}{\includegraphics[width=5cm]{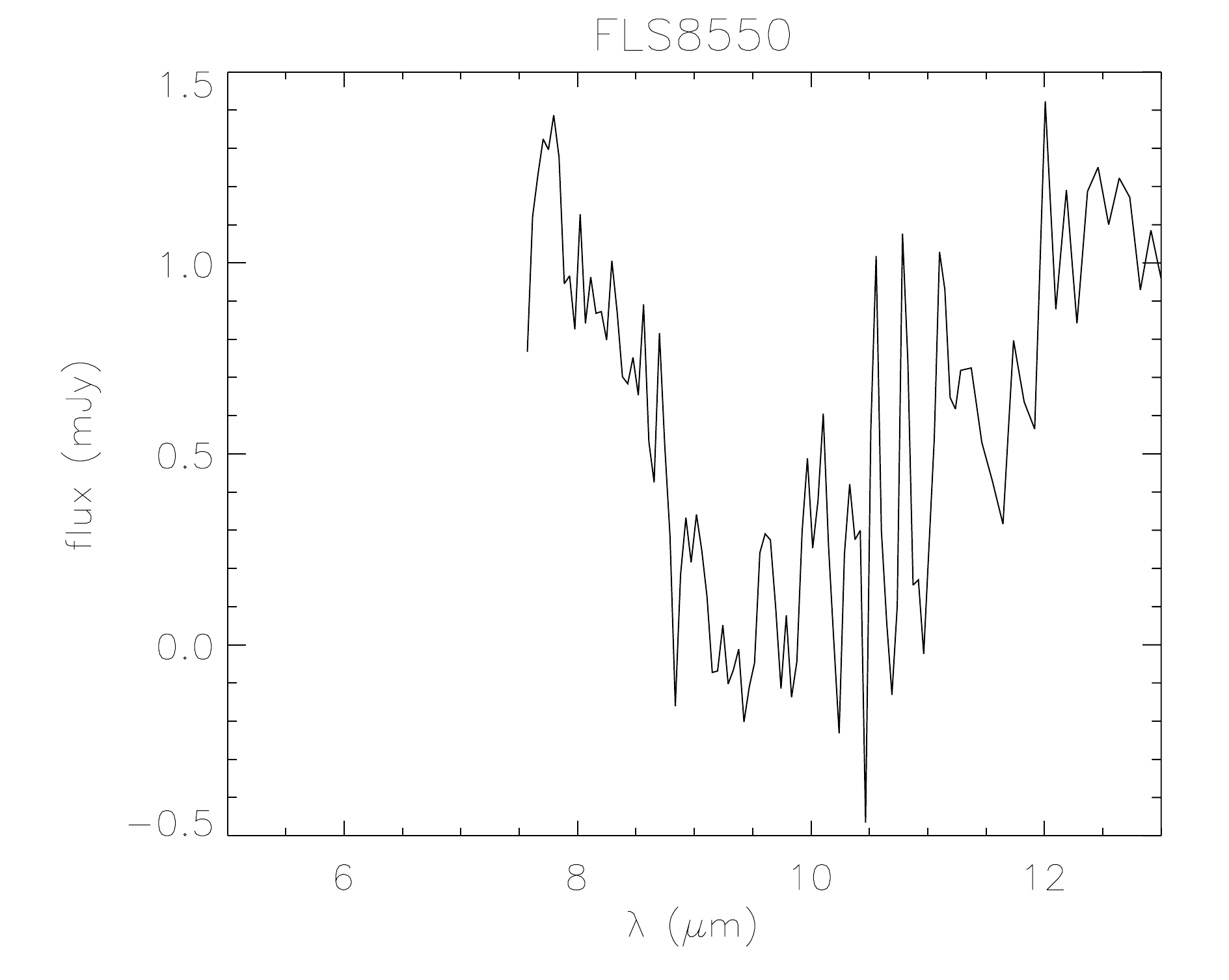}}\hfill \\
\caption{IRS spectra of the GOODS and FLS sample.}
 \label{irs}
 \end{figure*}

Since the initial sample selection is made at infrared wavelengths, we 
expect to find  a significant number of non-AGN sources, which owe their infrared
emission to dust heated by star-formation processes. To minimize 
 the contamination by pure star-forming systems, we exclude
sources with significant PAH emission at 6.2 or 11.3 or $\rm \mu m$ (see
Fig.\,\ref{irs}), since it has been demonstrated that AGN lack strong PAH
features \citep[e.g.][]{Genzel1998,Hernan2009}. 
  We  define as PAH-poor sources (mostly AGN dominated) 
  those that have  EW(6.2$\rm \mu m$)$<$0.3$\rm \mu m$, 
   or  EW(11.3$\rm \mu m$)$<$0.3$\rm \mu m$
  where the 6.2 $\rm\AA$ wavelength is not covered.   

There are 220 sources with IRS spectroscopy in the FLS sample, and we end up
with 20 high-$\tau$ AGN in FLS using the method described above. 
  Among these, there are six sources with \chandra  observations available.
  Five of them have no obvious PAH features (see Fig. \ref{irs} and Table \ref{irhigh}), 
   while one (FLS-283) has $\rm EW(11.3\mu m)$=0.68\micron and is therefore 
    not included in our sample.    Pope et al.
(in preparation) have consistently reduced and analyzed all publicly available
IRS spectra in both GOODS-N and ECDFS fields. For this paper, we search through
the 150 GOODS IRS spectra database to find AGN with high $\tau_{9.7}$, selecting
  15 sources. Eight of these sources are excluded because they 
 have prominent PAH features with EW larger than 0.3\micron (see Table \ref{rejected}).  
 The properties of the 12 sources in the high-redshift sample (seven GOODS 
  and five FLS sources) are given in Table \ref{irhigh},
  while their IRS spectra are shown in Fig. \ref{irs}.

\subsection{Comparison between the high and low redshift samples}
It is important to highlight the differences between the selection 
 at low and high redshift. In  the former, all sources are bona-fide AGN
according to optical spectroscopy.  Many sources would be classified as 
   composite AGN/star-forming systems, 
 on the basis of the  EW of the 6.2$\rm \mu m$ PAH feature. 
     
 Since there is no optical spectroscopy available for many sources in 
 the high redshift  sample, the AGN
classification is based on the absence of strong PAH features. 
 As discussed in section 4.3.2, our AGN selection criterion is shown to be effective for at least 
  eight out of twelve sources
   based on their high X-ray luminosities, and optical spectra and 
    most probably  all sources based on the measured 
   MIR excess in their SED fitting.  
 Therefore, the local and the high-redshift samples have
  different relative contributions to the star-formation 
   in the total IR energy budget.  Four sources in our 
   local sample (Mrk938 NGC3079, Mrk 266, Arp220)
   would not satisfy the $\rm EW(6.2\mu m)<0.3$\micron selection criterion imposed 
    in the high redshift sample.

\section{Analysis}

\subsection{The local sample}

\subsubsection{X-ray observations}
X-ray photometric observations exist for all sources in the local AGN sample.
For one source (NGC\,1125) only {\it Swift}-BAT observations
are     available at very high energies \citep[$\rm >15\,keV$;][]{Cusumano2010}, and no
X-ray spectrum can be derived. For another source (IRAS\,08572+3915) there are
{\it Chandra} observations available \citep{Iwasawa2009}, but unfortunately
they are not sensitive enough to allow the derivation of the X-ray column
through X-ray spectroscopy. For the remaining nine sources, we compile the X-ray
spectroscopic results available in the literature in Table\,\ref{xlocal}.

\subsubsection{X-ray spectra}
The classification of a source as Compton-thick is based on one of the
following criteria \citep[see][]{Georgantopoulos2009}: 
\begin{enumerate}[(a)]
\item{The most reliable criterion is the detection of the absorption turnover
      at high energies. These are the transmission-dominated Compton-thick
      sources with relatively `mild' column densities of a couple of times
      $\rm 10^{24}\,cm^{-2}$. In these sources, we can directly view the obscured
      component through the torus, even at energies below 10\,keV
      \citep[e.g. NGC3079;][]{Akylas2009}}.
\item{The detection of a high ($\rm \sim 1\,keV$) equivalent-width (EW) Fe
      line.}
\item{A `hard' spectrum with $\Gamma \sim 1$ or flatter. This is considered to
      be the signature of a reflection dominated Compton-thick source, where
      the X-ray emission comes solely from reflection from the backside of the
      torus \citep[see e.g.][]{Murphy2009}.}
\end{enumerate}
Six of the nine sources for which X-ray spectra can be derived probably 
host a Compton-thick nucleus based on one of the above criteria (see
Table\,\ref{xlocal}). Out of these, UGC\,5101 is probably the most debatable
case: in a study of ULIRGs with large amounts of molecular gas,
\citet{Georgantopoulos2010} detect a flat spectrum ($\Gamma \sim1.2$) using
{\it Chandra}, together with a high EW ($\sim$3.6 keV) FeK$\alpha$ line, in
agreement with \citet{Ptak2003}. The combination of a flat spectrum with a high
EW FeK$\alpha$ line suggests a Compton-thick AGN. \citet{Imanishi2003} analyzed
both the {\it XMM-Newton} and the {\it Chandra} data and found a considerably
lower EW ($\sim$400\,eV). \citet{Gonzalez2009}, using {\it Chandra},
claim the direct detection of a mildly Compton-thick AGN with a column density
of $\rm N_H \approx 1.4 \pm0.4\times 10^{24}$\,\cunits. 
 Finally, \citet{Brightman2010} report  a column density of 
  $10^{23}$ \cunits by re-analyzing the available \xmm data. We note that the same authors 
   report a much lower column density ($2\times 10^{23}$ \cunits) for Mrk266 (NGC5256)
    than the one derived by \citet{Guainazzi2005}.  NGC\,7582 is one of the
few known `changing-look' AGN, i.e. sources with past X-ray observations in
both a Compton-thin and a reflection-dominated state \citep{Bianchi2009}. The
three high-$\tau$ AGN that are definitely not Compton-thick are Mrk\,938, Mrk\,273, and
NGC\,7172. These have column densities between
$\rm N_H=8\times 10^{22}$\,\cunits and $4\times 10^{23}$\,\cunits. For Mrk\,273,
\citet{Iwasawa2011} note that the X-ray emission may come from a different
location than the IR nucleus. They also point out that the IR nucleus may be
associated with a Compton-thick source, although the evidence for this remains
weak. 

\subsection{Sources at high redshift}

\subsubsection{The X-ray observations}

The FLS has few (and shallow; $\sim 30$\,ksec) X-ray data points, which are presented in
\citet{Bauer2010}. Their X-ray observations targeted primarily ULIRGs. 
 There are X-ray observations available (all of them yielding upper
limits) for only five out of the 20 high-$\tau$ AGN in FLS. However,
the GOODS  survey for the {\it Chandra} Deep Fields has the deepest X-ray images to date.
The CDF-N was observed for 2\,Ms \citep[see][]{Alexander2003}. We examine
the 2\,Ms CDF-N X-ray images and find that all  sources are
detected in the total (0.3-8\,keV) band.
Three of the sources are in the catalog of \citet{Alexander2003},
while the other one  is marginally below the detection threshold of these authors. 
 The CDF-S was originally observed for 2\,Ms \citep{Luo2008}, while  the
field was observed for an additional 2\,Ms. Here, we use the entire set of 
  4\,Ms images. In the CDF-S, two sources (out of the three high-$\tau$ sources
in this field) are detected in the total (0.3-8\,keV) band.

\subsubsection{X-ray spectra}

 Among the X-ray detections, we can derive low-quality X-ray spectra in  five cases but  
 in only three we can constrain both the photon-index and the hydrogen column density.  
  We use
XSPEC \citep{Arnaud1996} for the spectral fitting. We use the C-statistic
\citep{Cash1979}, leaving the data ungrouped. The errors correspond to the 90\% 
confidence level. We fit a simple power-law with photoelectric
absorption from cold material in all cases. First, leave both the 
photon index and column density as free parameters for the three sources with 
the highest quality photon statistics. The results are given in Table\,\ref{xhigh}.
There is no clear evidence of a Compton-thick source. GN\_IRS-30 presents a
very flat spectrum, with $\Gamma=0.54$, but very large errors. To
obtain more stringent constraints on the absorbing column density, we fix the
photon-index to the commonly observed value of $\Gamma=1.8$ \citep{Dadina2008}. 
We see that the spectra of the sources GN\_IRS-19 and GN\_IRS-30 are
heavily obscured  with column densities $\sim10^{23}$ \cunits.
 
\subsubsection{Spectral energy distributions}

 To estimate the total infrared emission of the sources in the high
redshift sample, and assess the possibility that they host an AGN, we fit their
broad-band spectra with combinations of starburst and AGN templates. The
starburst templates we use come from the SWIRE template library
\footnote{{\tiny www.iasf-milano.inaf.it/\~\,polletta/templates/swire\_templates.html}}
\citep{Polletta2007}, which is a compilation of oberved SED of nearby galaxies,
and from the sample of \citet{Chary2001}.  The AGN templates
we use come from \citet*{Silva2004}, who combine nuclear SEDs of Seyferts with
a range of absorption columns. We also construct a sample of AGN templates
using the type-1 QSO SED from \citet{Richards2006} and applying dust absorption
following \citet{Rosenthal2000},  to account for the $\rm 9.7\,\mu m$
absorption feature.

The photometry we use for the GOODS-N and GOODS-S sources comes from the IRAC
(3.6, 4.5, 5.8, and $\rm 8.5\,\mu m$) public GOODS data release and the SIMPLE
survey \citep[{\it Spitzer} IRAC/MUSYC Public Legacy in ECDF-S;][]{Damen2010}
respectively. MIPS photometry at 24 and $\rm 70\,\mu m$, comes from the FIDEL
survey
\citep[Far-Infrared Deep Extragalactic Legacy survey; see][]{Magnelli2009} and
the $\rm 850\,\mu m$ photometry in the GOODS-N from \citet{Pope2005}. The FLS
sources have IRAC and MIPS ($\rm 70\,\mu m$) photometric values published in
\citet{Sajina2007}, while their MIPS ($\rm 24\,\mu m$) and 1.2\,mm fluxes are
published in \citet{Sajina2008}. In the IRS wavelength range, we bin the data
every five datapoints and average their flux values, which gives us 30
photometric data points per source.

\begin{figure*}
\rotatebox{0}{\includegraphics[width=5cm]{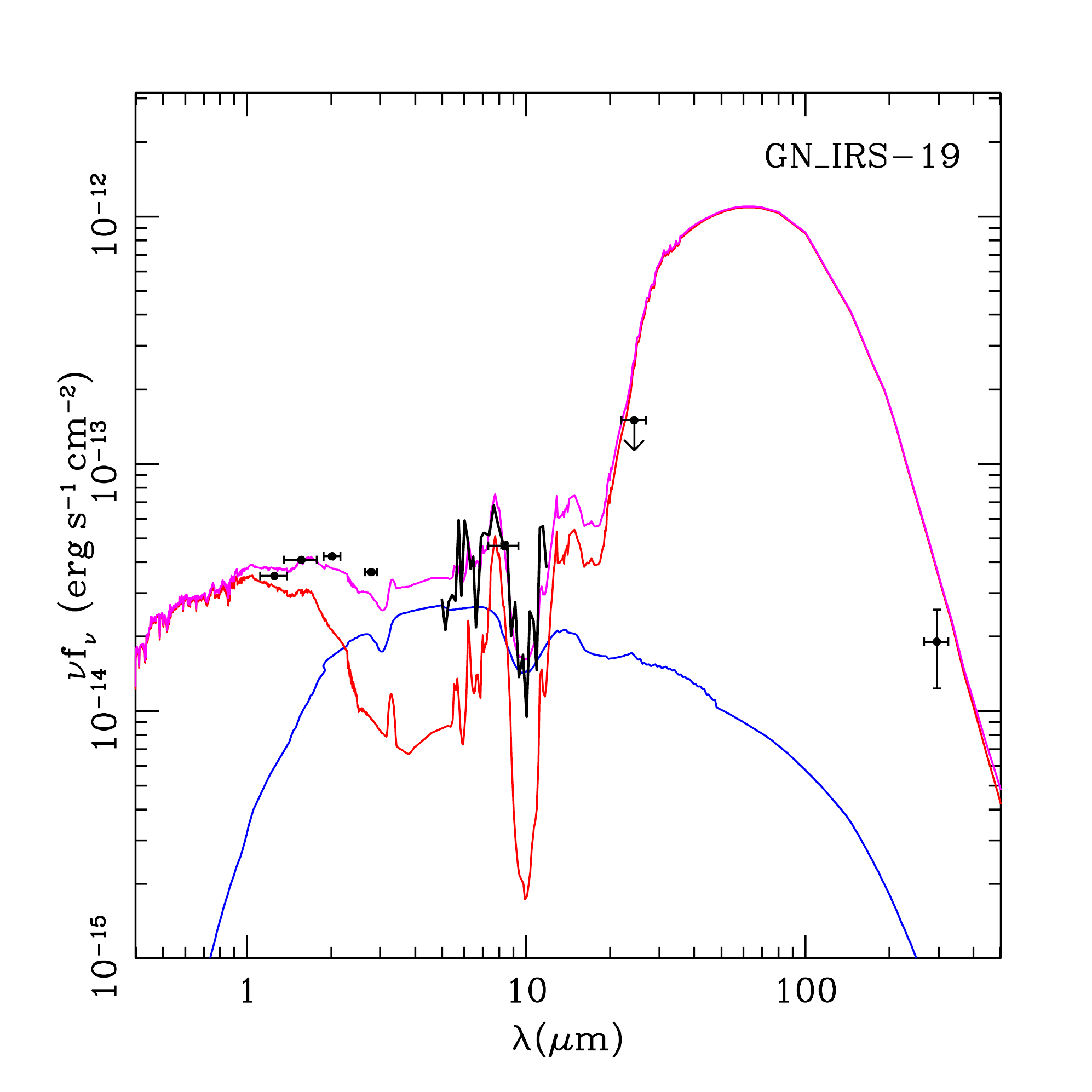}}
\rotatebox{0}{\includegraphics[width=5cm]{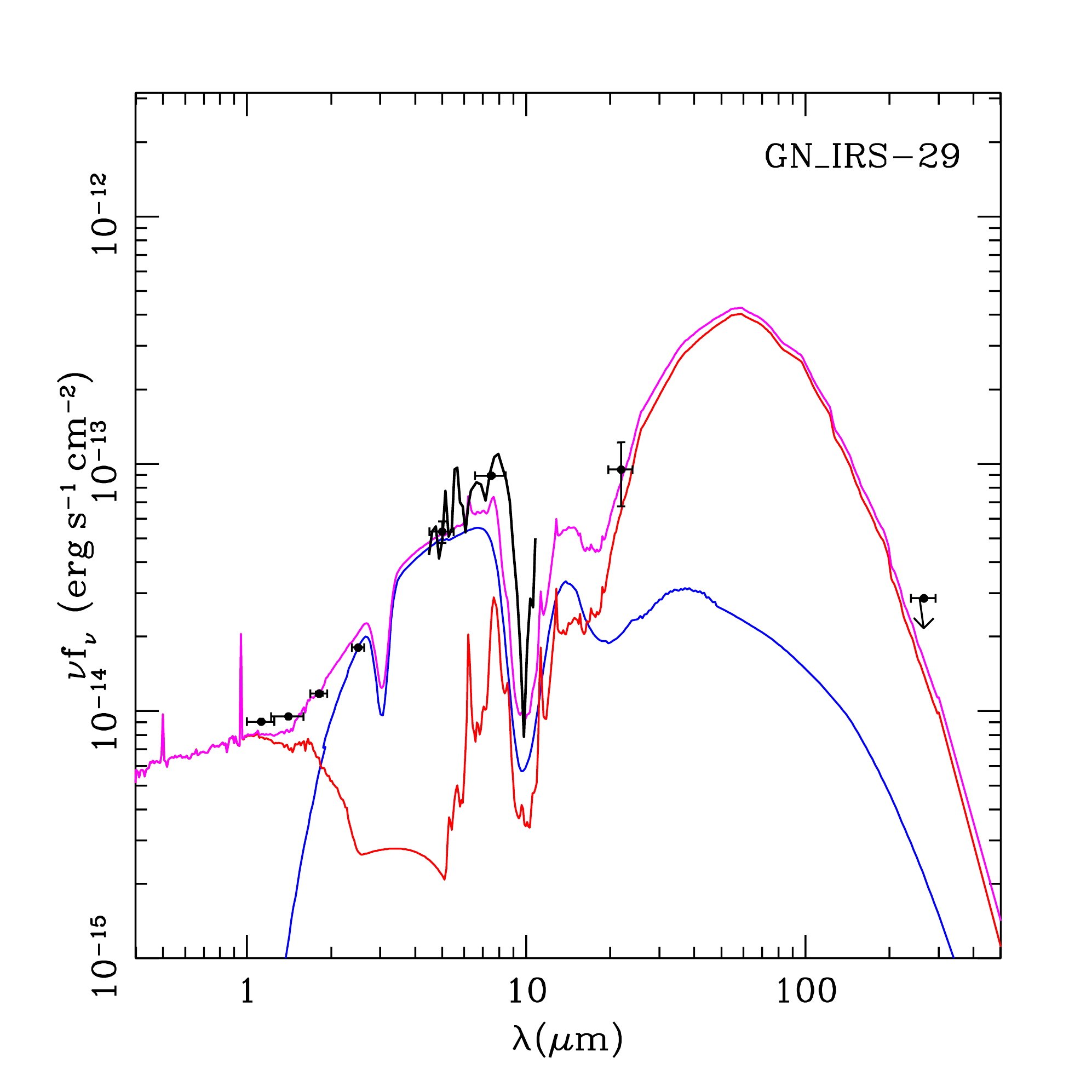}}
\rotatebox{0}{\includegraphics[width=5cm]{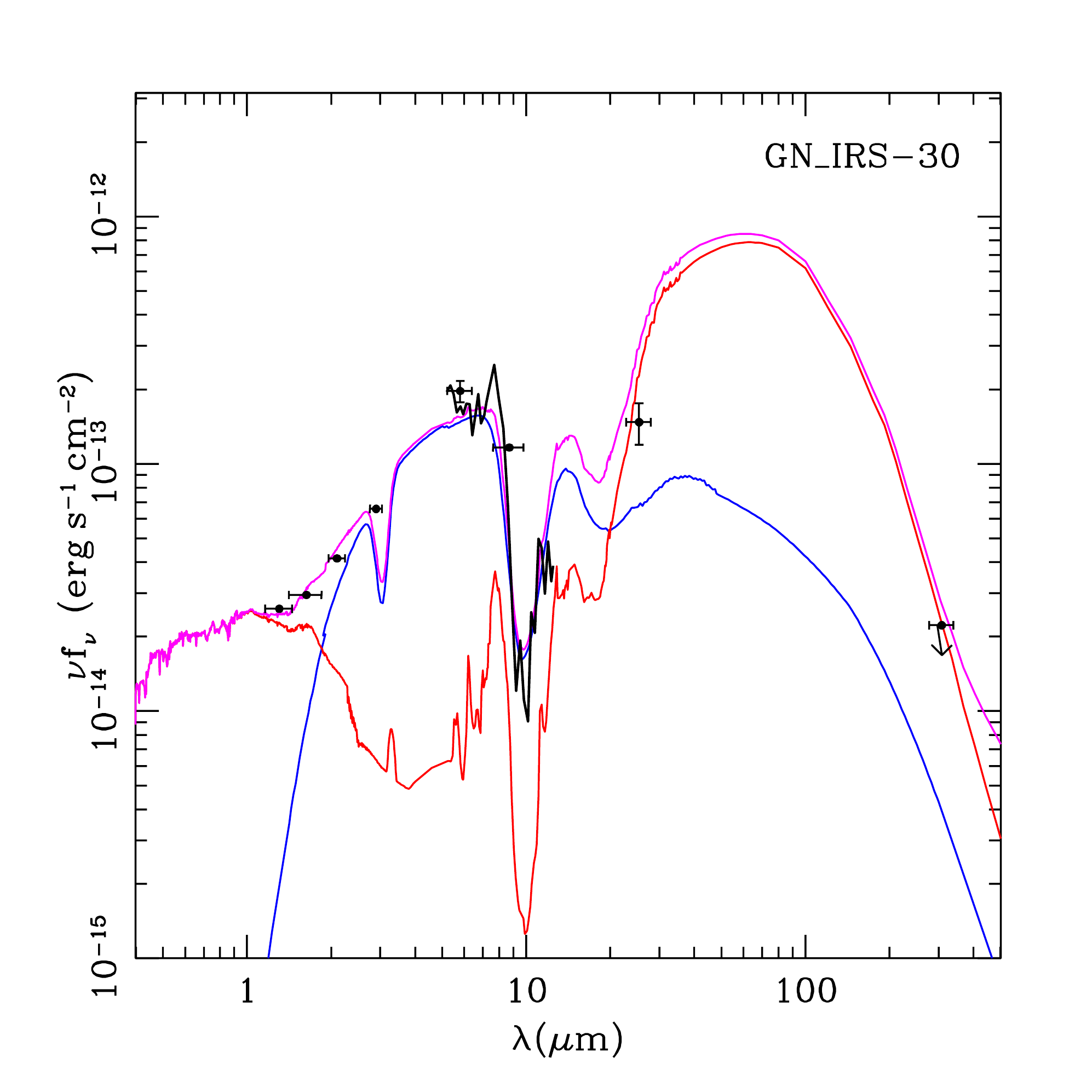}}\hfill \\
\rotatebox{0}{\includegraphics[width=5cm]{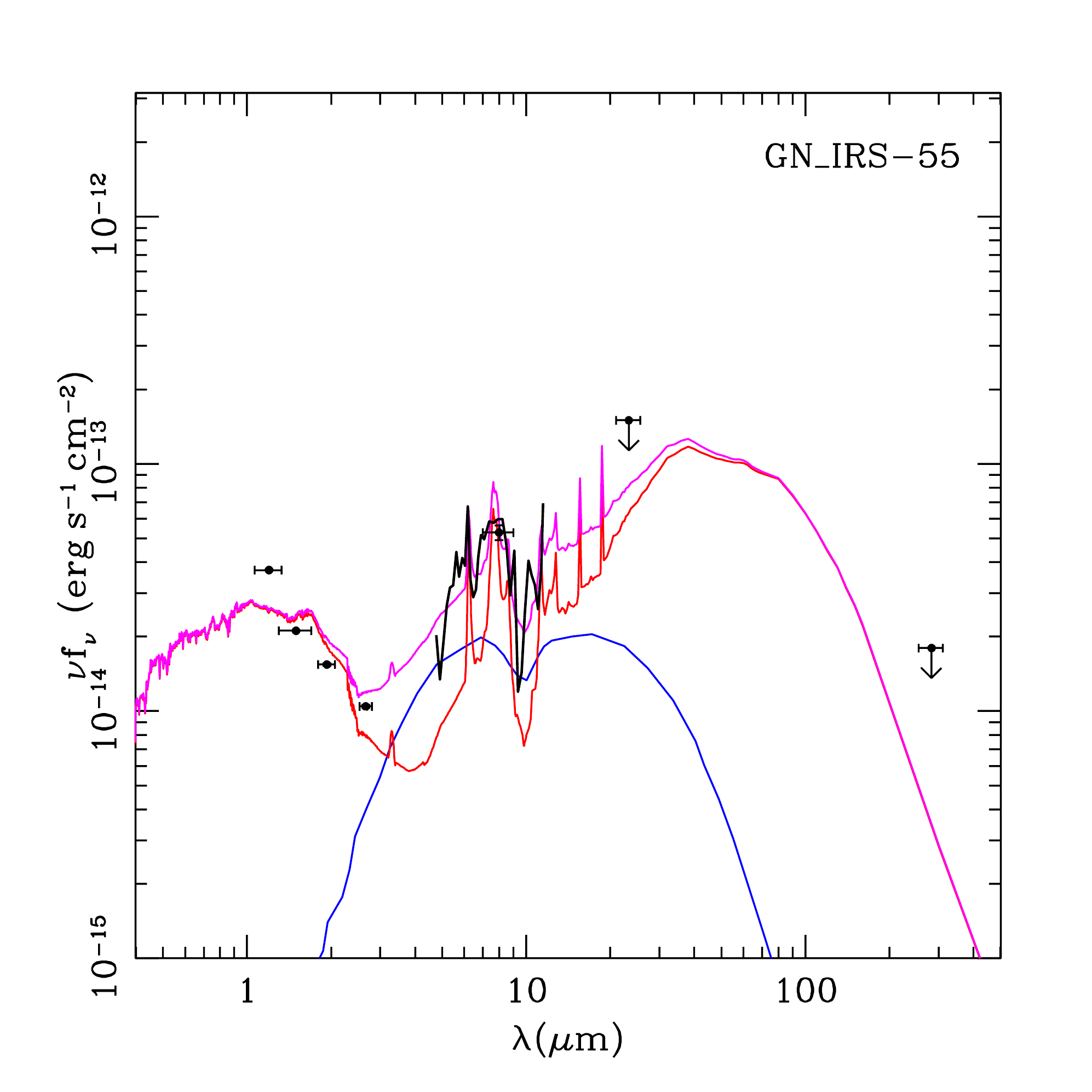}}
\rotatebox{0}{\includegraphics[width=5cm]{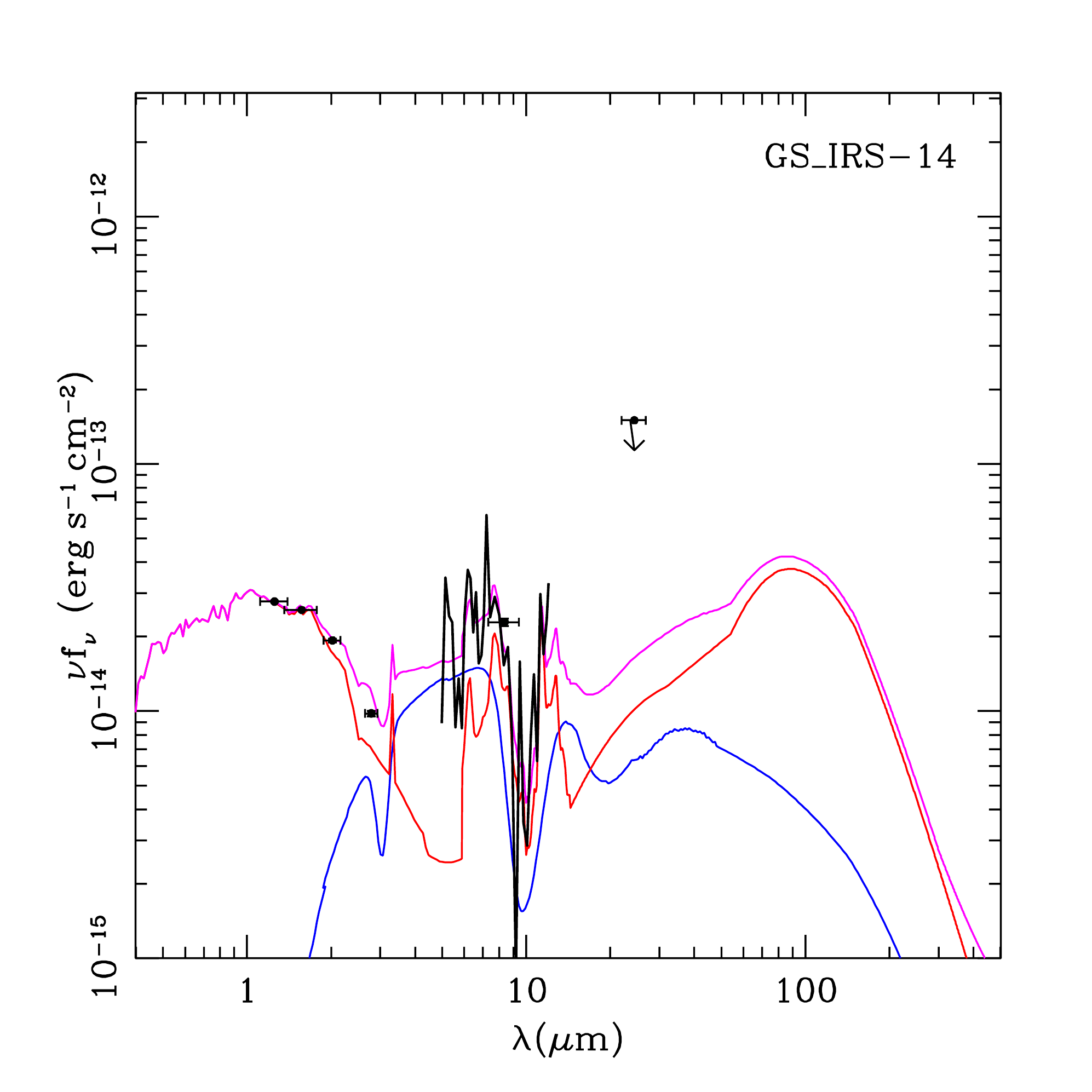}}
\rotatebox{0}{\includegraphics[width=5cm]{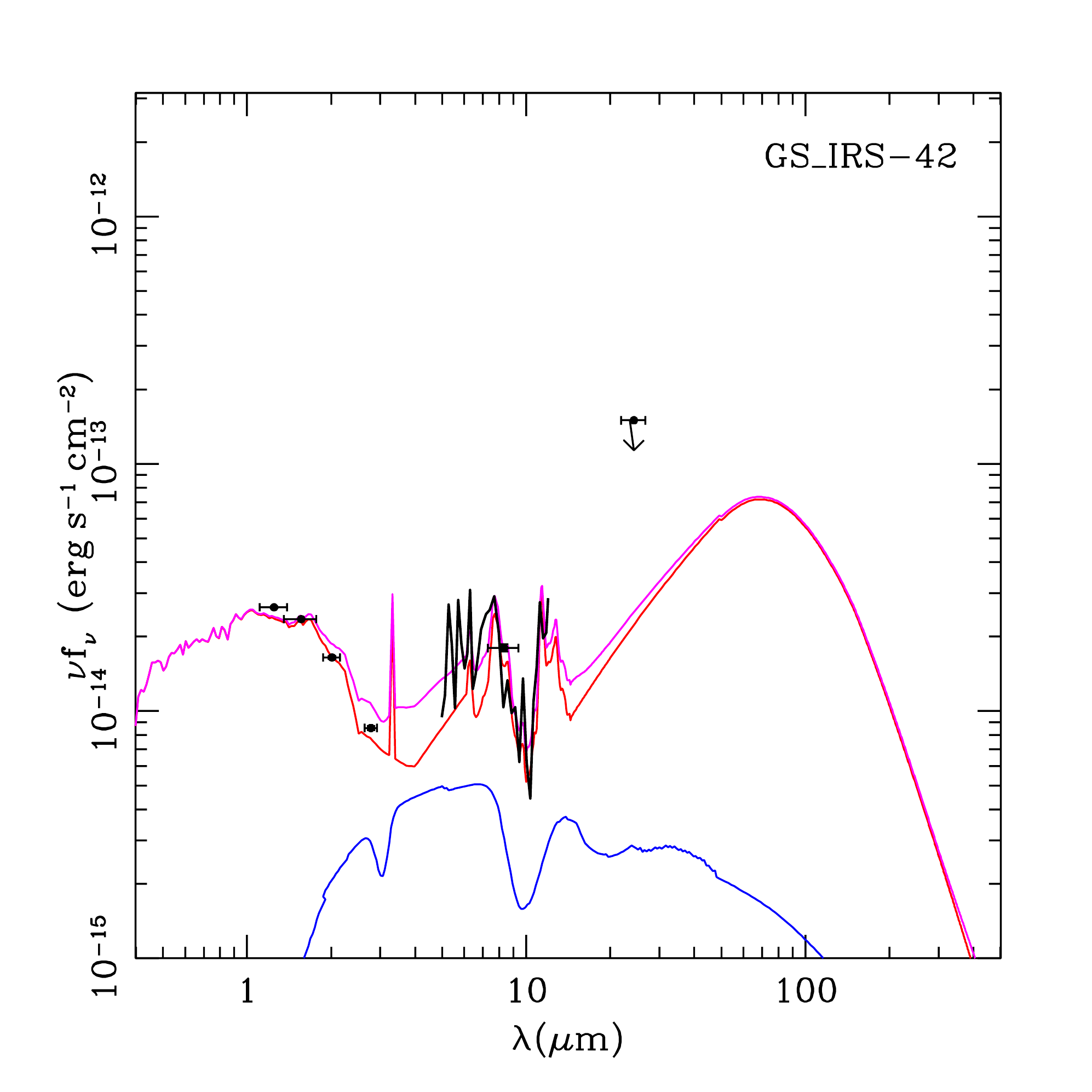}}\hfill \\
\rotatebox{0}{\includegraphics[width=5cm]{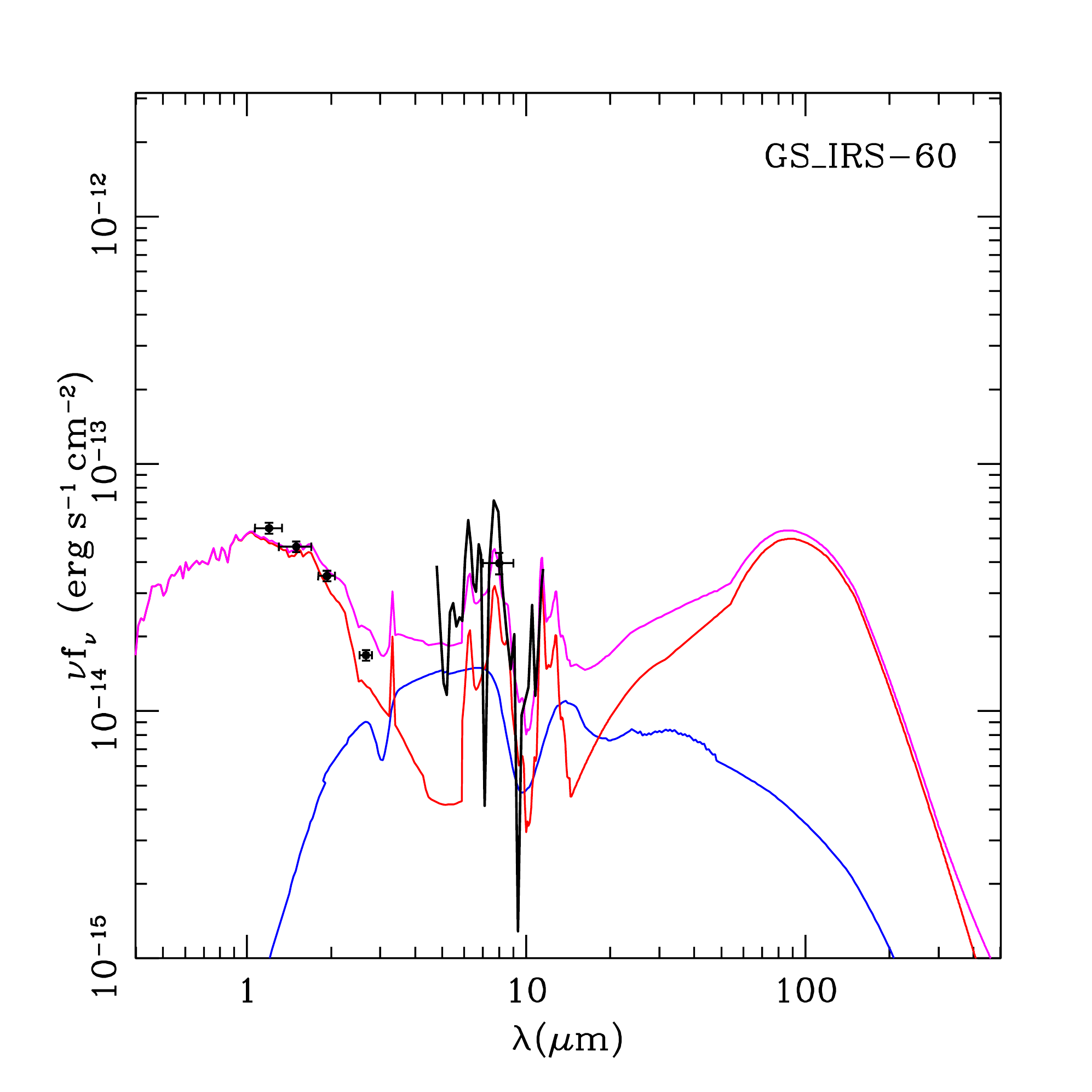}}
\rotatebox{0}{\includegraphics[width=5cm]{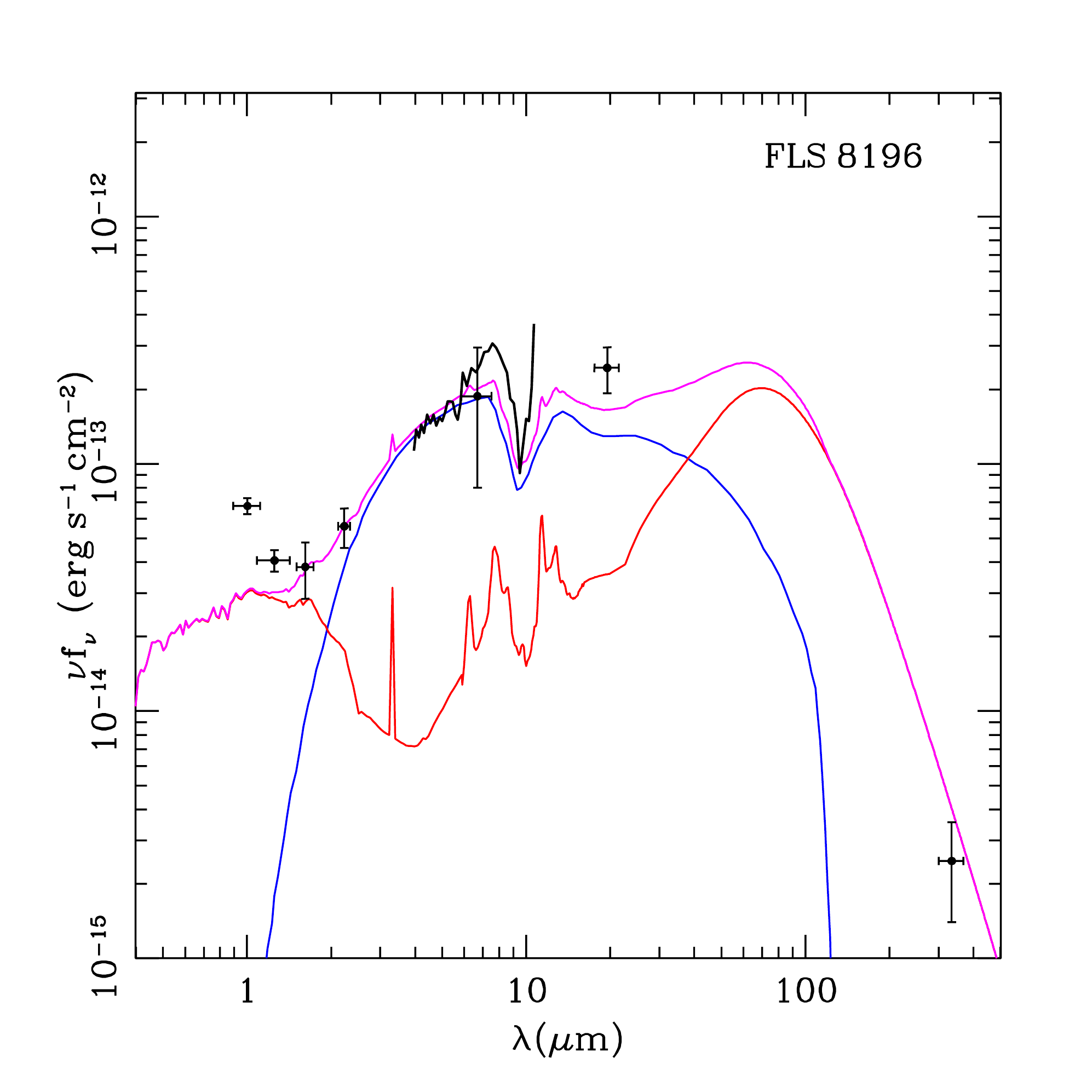}}
\rotatebox{0}{\includegraphics[width=5cm]{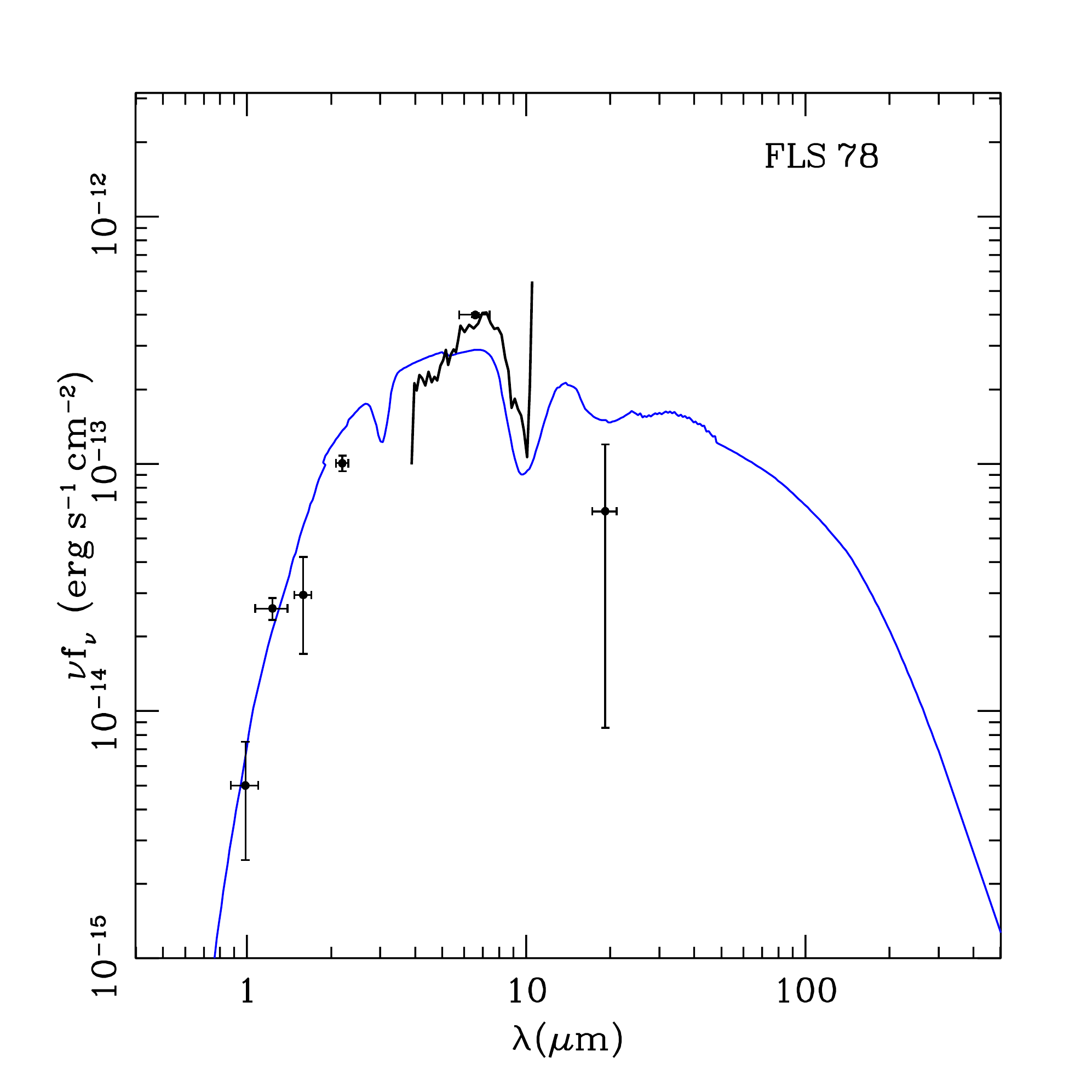}}\hfill \\
\rotatebox{0}{\includegraphics[width=5cm]{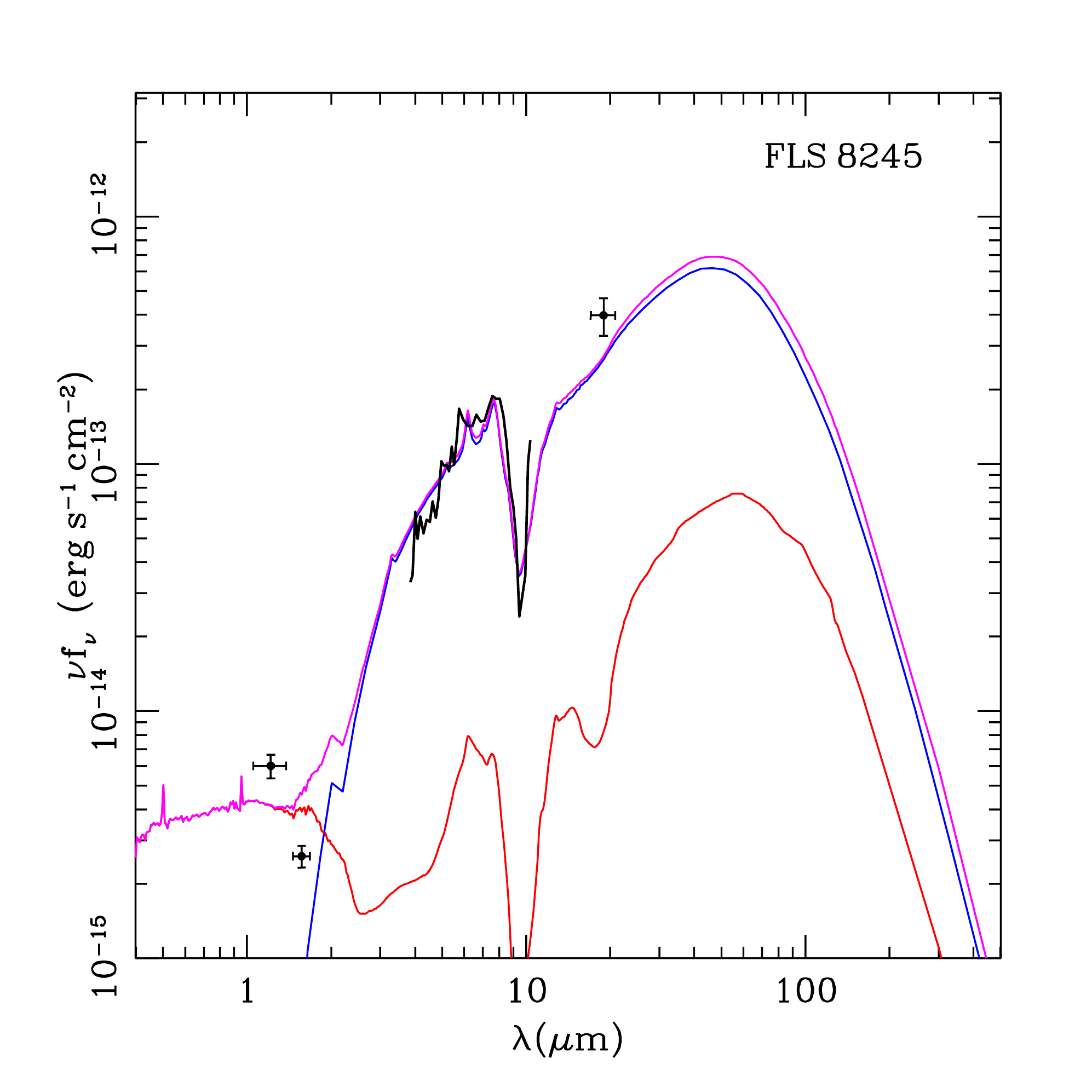}}  
\rotatebox{0}{\includegraphics[width=5cm]{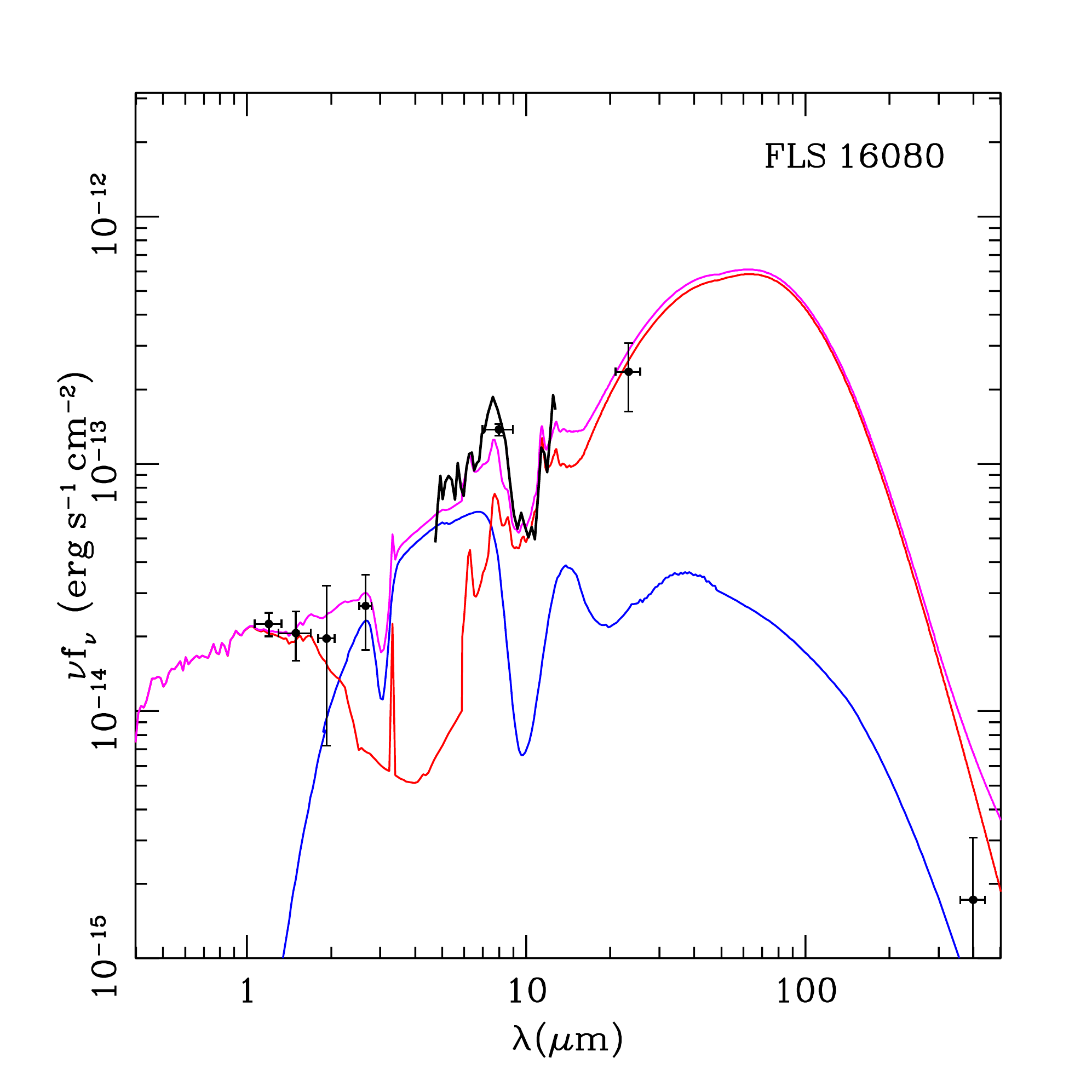}}
\rotatebox{0}{\includegraphics[width=5cm]{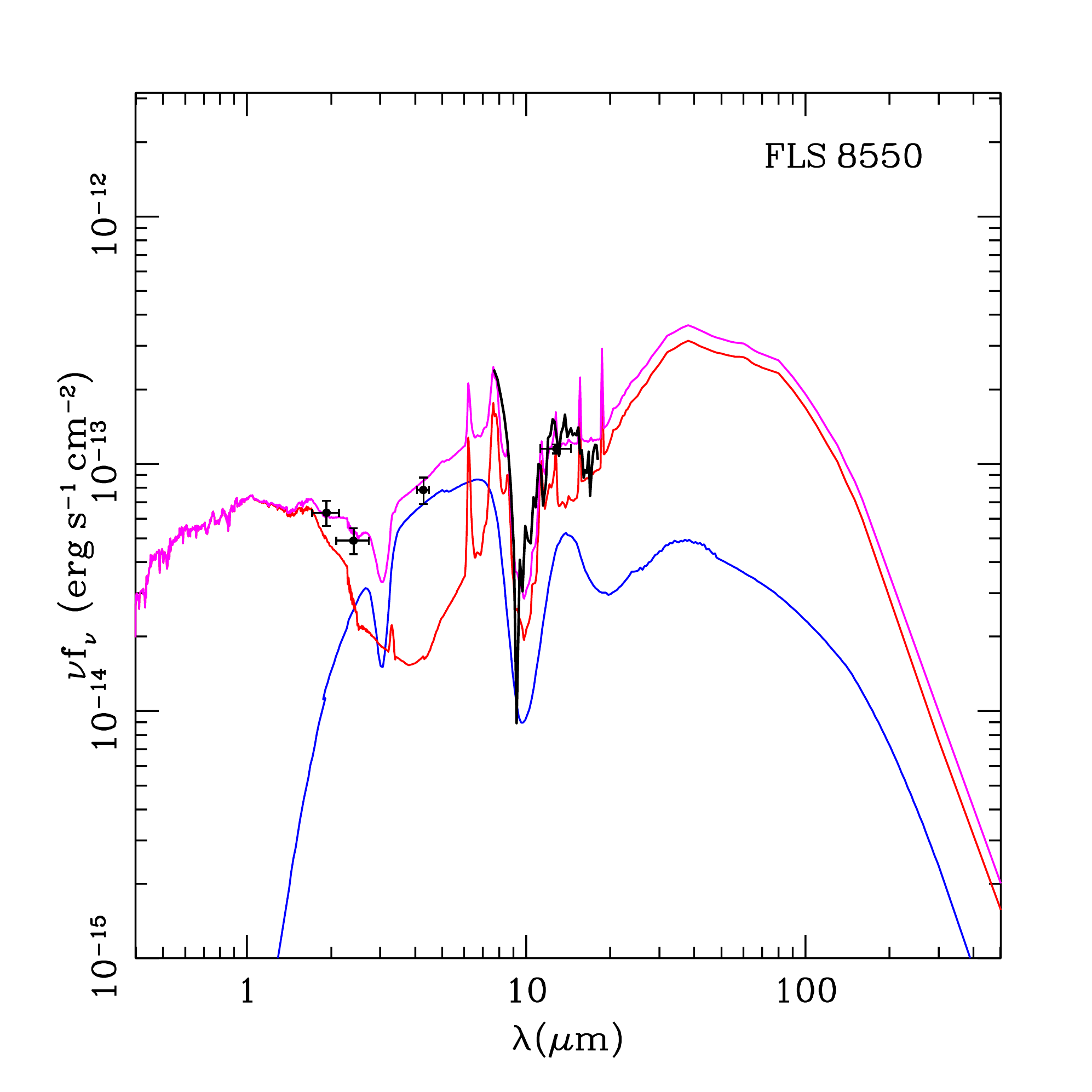}}\hfill \\
\caption{Mid-IR rest-frame SEDs  of the GOODS and FLS sample. The blue and red line
 denote the torus and star-forming contribution while the purple line denotes the sum of the two.}
 \label{sed}
 \end{figure*}

We use a $\chi^2$ minimisation method to select the optimum combination of a
starburst and an AGN template to fit our data. The results can be seen in
Fig.\,\ref{sed}. We calculate the $\rm 8-1000\,\mu m$ luminosity by
integrating the best-fit SED; the results are
listed in Table\,\ref{irhigh}. We note that six of the GOODS-N sources
have their infrared luminosities published in \citet{MurphyIR2009}, where a
similar procedure was used in their calculation. Our $L_{IR}$ values
agree within a factor of 2 with the $L_{IR}$ values of \citet{MurphyIR2009}.
Most of the sources of Table\,\ref{irhigh} require both a star-forming and an
AGN template to fit the photometric points. The spectral decomposition 
(Fig. \ref{sed})  suggests that the bulk of the mid-IR emission is produced by the AGN component. 
 The exception is GS\_IRS-42. In the case of this source, an
 F-test gives only an 8\% probability that the AGN component is needed.

\begin{table*}
\centering
\caption{High-$\tau$ $\rm 9.7\,\mu m$ AGN at higher redshift}
\label{irhigh}
\begin{tabular}{ccccccccccccc}
\hline
Name       &  $\alpha$   & $\delta$    & z         & $S_8$ & $\tau$ & PAH EW   & $\rm log[L_{6}]$ & $\rm log[L_{IR}]$ & AGN & Comment   & AGN      \\
 (1)       &    (2)      & (3)         & (4)       &   (5)  & (6)            & (7)                     & (8)               & (9)         & (10)   & (11)  & (12) \\
\hline
\multicolumn{7}{l}{CDF-N} \\ \hline
%GN\_IRS-16 & 12:36:37.0  & +62:08:52   & 2.02$^2$  & 0.015 & 3.97   &    $<$0.23&  11.27                  &  12.29$^1$        & 0.75 &  a,d,e \\
GN\_IRS-19 & 12:35:55.1  & +62:09:01   & 1.875$^1$ & 0.097 & 2.43   &   $<$0.18  &  11.37                  &  12.62            &   0.75 &  a,b,d,e & Y       \\
%GN\_IRS-21 & 12:36:18.3  & +62:15:50   & 2.00$^2$  & 0.020 & 1.96   &   0.27  &  11.19                  &  12.60            &   0.73 & a,e \\
GN\_IRS-29 & 12:36:56.5  & +62:19:37   & 2.2$^2$   & 0.048 & 2.54   &   $<$0.11   & 11.83                  &  12.62$^1$        & 0.91& a,d,e & Y \\
GN\_IRS-30 & 12:37:26.5  & +62:20:26   & 1.76$^2$  & 0.175 & 3.6    &    $<$0.07   & 11.92                  &  12.74$^1$        &  0.96 & a,d,e & Y \\
GN\_IRS-55 & 12:36:46.7  & +62:14:46   & 2.004$^3$ & 0.027 & 1.8    &   $<$0.09    & 11.49                  &  12.40            & 0.58 & a,b,d,e   & Y   \\
\hline
\multicolumn{7}{l}{CDF-S} \\ \hline 
GS\_IRS-14 & 03:32:14.5 & -27:52:33 & 1.87$^2$  & 0.026 & 2.22   &    $<$0.20    & 11.16                  &  12.66$^1$        & 0.67 & d,e & p \\
GS\_IRS-42 & 03:32:25.9 & -27:47:51 & 1.88$^2$  & 0.02 & 1.78   &    $<$0.28   & 10.99                  &  12.25$^1$        &  0.32 & a,d    & Y   \\ 
GS\_IRS-60 & 03:32:40.0 & -27:47:55  & 2.0$^2$  & 0.045  & 1.35   &   $<$0.17    & 11.78                 &    13.25                  &   0.57&	a,d,e & Y \\
\hline
\multicolumn{7}{l}{FLS} \\ \hline
FLS-8196       & 17:15:10.2  & +60:09:54 & 2.59$^2$  & 12.5    & 1.32   &    $<$0.05    & 12.50                  &  13.40            &  0.74  & d,e, f  & Y \\
FLS-78         & 17:15:38.2  & +59:25:40 & 2.65$^2$     &12.7  & 1.49   &   $<$0.04    &  12.71                  &  13.51            & 1 & d,e &  p \\
FLS-8245       & 17:15:36.3  & +59:36:14 & 2.70$^2$    & 24.   & 2.13   &     $<$0.05    & 12.38                  &   13.25            & 0.98 & d,e & p                \\
FLS-16080      & 17:18:44.8  & +60:01:15 & 2.01$^2$   &71.   & 2.17   &      $<$0.08  &  11.76                  &  12.85            &  0.71 & c,d,e & Y \\
FLS-8550       & 17:18:14.6  & +59:56:05 & 0.87$^2$     & 255.  & 2.74   &      $<$0.27  & 11.10                  &  12.09$^1$        & 0.70 &d,e & p \\ 
\hline
\end{tabular}
\begin{list}{}{}
\item The columns are: (1) Name
                       (2),(3) Equatorial {\it Spitzer} coordinates
                       (4) Redshift: $^1$  \citet{Chapman2005} ;
                                     $^2$ IRS from \citet{Pope2008} or \citet{MurphyIR2009}
                                      in the case of GOODS, 
                                     or \citet{Sajina2007} in the case of FLS..
                                     $^3$ \citet{Barger2008}.
                       (5) 8$\mu m$ flux in units of $\rm \mu Jy $. 
                       (6) optical depth at 9.7$\mu$m.
                       (7) PAH equivalent-width at 6.2 $\rm \mu m$ 
                        in units of \micron (except in the case of  FLS-8550
                         where the 11.3 \micron EW is quoted).
                        (8) Logarithm of the $\nu L_\nu$ monochromatic
                           luminosity at $\rm 6\,\mu m$ in units of solar
                           luminosity as derived by IRS spectroscopy.
                       (9) Logarithm of the 8-1000\,$\mu$m IR luminosity
                           in units of solar luminosity as derived from the SED
                           fitting.  $^1$:
                           No photometry at rest-frame wavelengths
                           $\rm >30\,\mu m$ which constrains the far-IR
                           luminosity.
                       (10) Fraction of AGN contribution at 6$\rm \mu m$ according to the spectral decomposition.
                      (11) Evidence for the presence of an AGN: a) X-ray luminosity b) radio-emission c) broad optical line 
                           d) absence (or weakness)  of PAH feature e) SED f) broad-ish ($>500 \rm km~s^{-1}$) [OIII] line 
                           (12) AGN classification: (Y) secure on the basis of the X-ray luminosity or optical spectroscopy.
                            (p) probable on the basis of the mid-IR diagnostics i.e. low-EW PAHs and 
                             SED fitting.  
\end{list}
\end{table*}

\begin{table*}
\centering
\caption{X-ray properties of the high-redshift sample}
\label{xhigh}
\begin{tabular}{crrrr}
\hline \hline 
        &     & \multicolumn{2}{c}{$\rm \Gamma, N_H$ free}    & $\Gamma=1.8$        \\
\hline                               
Name        &  $L_X$ & $\Gamma$               & $\rm N_H$            & $\rm N_H$            \\
 (1)       & (2)                    & (3)                  & (4)  & (5)                 \\
\hline 
CDF-N   &    &            &                                &                                                 \\
\hline
%GN\_IRS-16  &  42.1      &  -                           & -         &  -                      \\
GN\_IRS-19$^a$  & 43.5 & $1.44^{+0.47}_{-0.44}$ & $12.1^{+6.9}_{-5.5}$ & $16.8^{+4.0}_{-3.8}$ \\
%GN\_IRS-21  & 42.4  & -                    &  -                      & $<$6.5               \\
GN\_IRS-29   & 42.3 & -                    & -                      & $<$ 7.9              \\
GN\_IRS-30$^b$ & 43.1 & $0.54^{+1.18}_{-0.49}$ & $<$17.5              & $17.0^{+16}_{-10}$   \\
GN\_IRS-55$^c$  & 42.3 & $2.15^{+2.6}_{-1.18}$  & $<$5.8             & $<$4                 \\
\hline 
CDFS-S &    &         &                                &                                                    \\
\hline
GS\_IRS-14 &        $<$42.9     & - & - & - \\
GS\_IRS-42$^d$ &  42.3     &     -           & -          & $<$1.3       \\
GS\_IRS-60      & 42.1   &    -        &  -  & -  \\
\hline   
FLS &                 &           &                          &                                                     \\
\hline
8196  & $<$43.8  &    -      &             -               &                -                    \\
78      &  $<44.1$  & -         &            -               &               -                 \\
8245   &  $<$43.8 &     -      &             -              &                -               \\
16080  & $<$43.7  &  -         &               -            &              -                \\
8550    &   $<$43.7 &    -    &         -                    &            -             \\                 
\hline \hline
\end{tabular}
\begin{list}{}{}
\item The columns are: (1) Name ($^a$), ($^b$), ($^c$) sources 44, 423, and 243 respectively in the catalogue 
 of  \citet{Alexander2003}; ($^d$) source 240 in \citet{Luo2008}.
                     (2) Logarithm of the X-ray Luminosity (2-10 keV) (or $3\sigma$ upper limits) uncorrected for absorption in units \lunits. 
                        Luminosities have been estimated from the X-ray spectral fits. Where there is no spectral fit,  
                         we assume $\Gamma=1.4$. 
                       (3) Photon index.
                       (4) Rest-frame column density in units of
                           $10^{22}$\,\cunits.
                       (5) Rest-frame column density in units of
                           $10^{22}$\,\cunits, where $\Gamma=1.8$.
\end{list}
\end{table*}

\subsubsection{$\rm L_X/L_6$ ratio}
 For faint sources such as those in the CDFs, it is difficult
to apply X-ray spectroscopy diagnostics owing to the limited photon statistics.
Alternatively, the $\rm L_X$ to $\rm L_{6\,\mu m}$ ratio could be used to
identify any highly obscured, Compton-thick AGN. One of the most reliable
proxies of the intrinsic power of an AGN is considered to be the mid-IR
$\rm 6\,\mu m$ luminosity \citep[e.g.][]{Lutz2004, Maiolino2007}. This
wavelength region can contain  significant emission from  the hot dust heated by the AGN, and thus
provides a reliable diagnostic of the AGN power. On the basis of this,
\citet{Alexander2008} and \citet{Goulding2011} claim that very low X-ray to
$L_{\rm 6\,\mu m}$ ratios, typically 30-40 times lower than those of unobscured
AGN, is indicative  of very strong attenuation at X-ray wavelengths and
thus of Compton-thick sources.
 
\begin{figure*}
\begin{center}
\includegraphics[width=12.0cm]{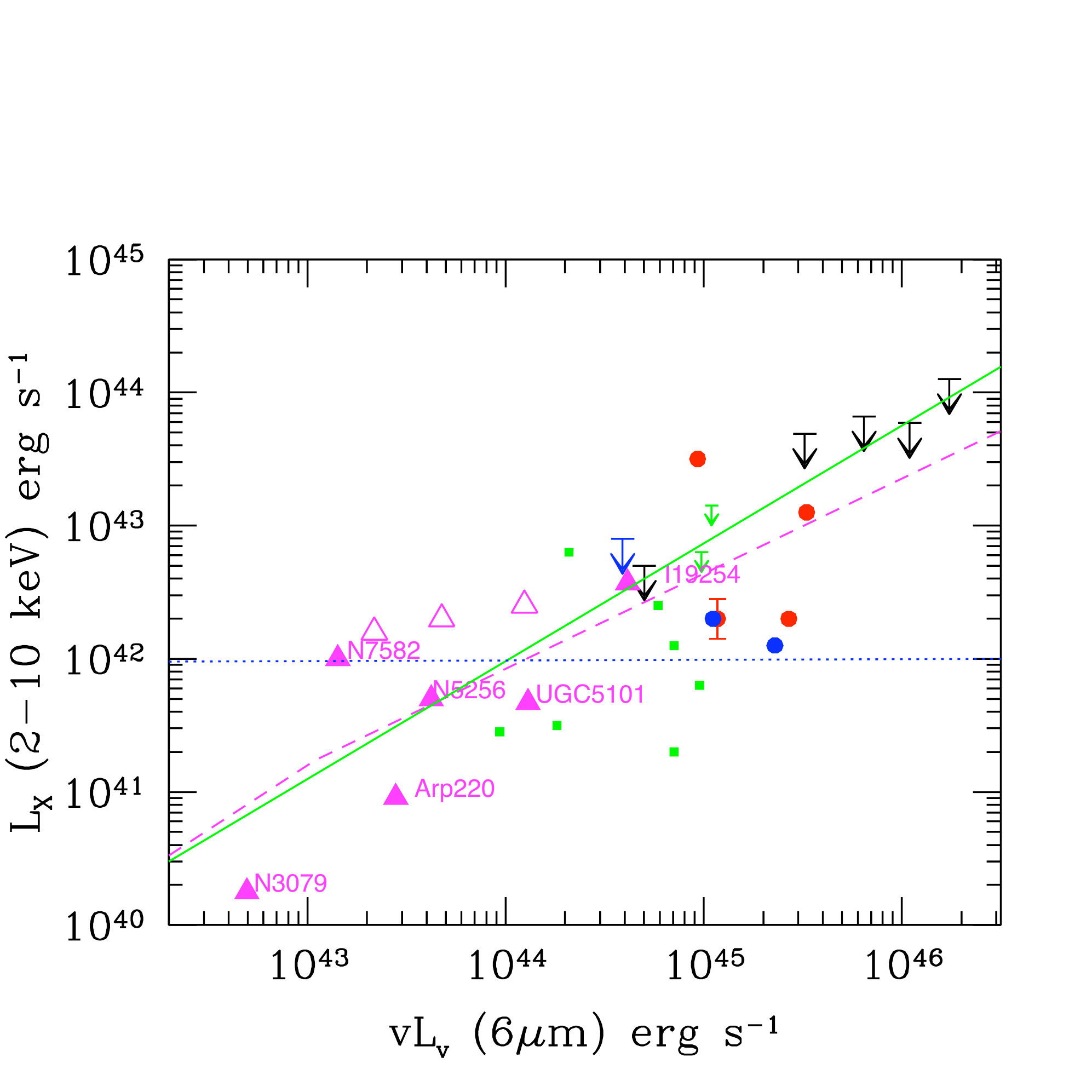} 
\caption{Rest-frame (uncorrected for absorption) $\rm L_X$ versus $\rm L_{6\,\mu m}$
         luminosity diagram. The filled triangles  correspond to the
         Compton-thick AGN in the $\rm 12\,\mu m$ sample, while the open
         triangles correspond to the three heavily obscured but not
         Compton-thick AGN in the same sample. The red (blue) filled circles or upper limits denote
         the high-$\tau$ sources in the CDF-N (CDF-S). The black upper
         limits denote the FLS points.  Finally, the green points (filled squares and  small upper limits) 
          denote the sources with considerable PAH emission that have not been included in our sample. 
           The dash line (magenta) denotes the low \lxl6 area that should be populated by Compton-thick AGN.
          This line is derived  from the average \lxl6 relation derived for 
           the COSMOS AGN \citep{Fiore2009} by scaling it down by a factor 0.03,  which 
            corresponds to the average flux of the reflection component 
             in the 2-10 keV band relative to the intrinsic power-law component in local 
              Compton-thick AGN \citep{Comastri2004}. 
             The solid line (green) 
         corresponds to the luminosity-dependent Compton-thick line based on
         the \citet{Maiolino2007} relation, scaled down by the same amount.
          Finally, the horizontal dotted line at $\rm L_X=10^{42}$ \lunits denotes the
           upper limit to X-ray emission due to star-forming processes \citep{Tzanavaris2006}.  
          }
\label{lxl6}
\end{center}
\end{figure*}

We present the 2-10\,keV luminosity, uncorrected for absorption, against the
monochromatic $\rm 6\,\mu m$ IR luminosity for both the $\rm 12\,\mu m$ sample
and the higher redshift sources in Fig.\,\ref{lxl6}. We define very roughly a
region where Compton-thick AGN reside, by scaling down the relation of
\citet{Fiore2009} for the AGN in the COSMOS fields by 0.03. This is very roughly the amount
of reflected emission in a type-2 AGN \citep{Comastri2004}. Moreover, we show
the scaled down $\rm L_X-L_{6 \mu m}$ relation, by the same amount,   
 of AGN in the local Universe found by \citet{Lutz2004}. Finally, we show the
luminosity-dependent line predicted by \citet{Maiolino2007}. 
 The luminosity error bars do not play a major role
relative to the uncertainties introduced by the use of different Compton-thick
lines. We present a typical error-bar for one source. This includes both the
photon statistics error as well as the uncertainty in the spectral model used
for the conversion from flux to luminosity. Therefore, the uncertainties in this 
diagnostic are quite large. 

Using as a guide our AGN in the local Universe for which we have reliable X-ray
spectroscopy, we see that there is one Compton-thick source (NGC\,7582) that 
lies above the local Compton-thick AGN regime. 
 Moving to our AGN at higher redshift, there are four GOODS sources that could be
classified as Compton-thick according to the $\rm L_X/L_{6 \mu m}$ diagnostic, i.e. 
  GN\_IRS-29, GN\_IRS-55, GS\_IRS-42, and GN\_IRS-60.   
Taking into account only the X-ray detections, the fraction of possible Compton-thick 
 sources would be 66\% (4/6). The X-ray upper limits are all situated unconveniently above 
 the Compton-thick regime. Under the extreme assumption that all upper limits 
  are either associated or not with Compton-thick sources, the fraction of Compton-thick 
   AGN could vary between 33\% and 83 \%.  

\subsubsection{Contamination by star-forming galaxies.}
  Our selection criterion (low EW PAH) is chosen to 
   minimise the number of star-forming galaxies (non-AGN)
    in our sample.  
  The EW upper limits of our FLS sources are consistent 
    with a high AGN contribution, close to 100\% according to Fig.5 of \citet{Sajina2007}.
        The absence of strong PAH features is unlikely to be an artefact of 
        low signal-to-noise ratio spectra. 
        There is no trend for a dependence of the EW   
 on the mid-IR flux (see Table 3). The two brightest
 FLS sources, with X-ray observations available,  are among the sources with the lowest EW. 
    Another possibility could be that the PAHs are possibly suppressed in star-forming 
     galaxies with large amounts of absorption \citep[e.g.][]{Zakamska2010}. 
     However, the region responsible for the silicate absorption is most probably located between the AGN torus 
      and the narrow-line-regionÊ\citep[e.g.][]{Netzer2007}.  The most solid evidence that 
      these silicates are mostly located close to the torus is that they are often seen 
       inÊemission in type-1 
      AGN and in absorption in type-2. In contrast,  most of the PAH 
      emission should originate from  larger  scales  \citep{Soifer2002}.   
      Furthermore, inspection of the strength of the
   6.2$\rm \mu m$ PAH feature  of the rejected $\tau>1$ sources in Table 5,
    shows that these display high 
       PAH EW (up to EW$\sim1.2$ $\rm \mu m$) despite there being  
      optically thick at 9.7$\rm \mu m$. 

 However, the most robust indicator of detection  
   of an AGN can be provided by the level of its X-ray emission. 
 Although a detection at X-ray wavelengths is generally sufficient to confirm 
 the presence of an AGN, the GOODS fields 
  probe extremely faint fluxes where a number of sources could be associated with normal galaxies (i.e.
without an AGN)  \citep{Georgakakis2007, Ptak2007}. These also have 
low $\rm L_X/L_{6 \mu m}$ ratios because their X-ray emission is intrinsically
weak (instead of absorbed). We set the AGN detection limit to $L_X =10^{42}$ \lunits as
no local pure star-forming galaxy has ever presented a 2-10 keV luminosity 
 above this threshold \citep[see][]{Tzanavaris2006}. 
 One of the most X-ray luminous star-forming galaxies is 
 NGC 3256, which has a 2-10 keV luminosity	of $2.5\times10^{41}$	\lunits,	
 but	displays no evidence of  AGN activity  \citep{Moran1999}.
 Therefore, for all our six sources with X-ray detections, we can
be confident that they are AGN, as they present X-ray luminosities 
 above $10^{42}$ \lunits.

  In agreement with the above classifications, 
 two of our sources (GN\_IRS-19 and GN\_IRS-55) have
additional evidence of an AGN via excess of radio emission
(Del Moro et al., in preparation). 
FLS-8196 presents an [OIII] line with a
FWHM of $\approx 800$ $\rm km~s^{-1}$ \citep{Sajina2008}, 
 which is typical of emission from the AGN narrow-line-region
       \citep{Zakamska2003}.  FLS-16080 displays a broad
$\rm CIV \lambda1550 \AA $ emission line, with a FWHM of
$\rm \approx 2500\,km\,s^{-1}$, and hence also hosts an
AGN \citep{Sajina2008}.
  The presence of a broad-line argues against 
  this source being Compton-thick. There are 
   cases of course of broad-line QSOs associated with Compton-thick AGN 
   but these cases are rare \citep{Braito2004}.

%%%%%%%%%%%%%%%%%%%%%%%%%%%%%%%%%%%%%%%%%%%%%%%%%%%%%%%%%

\begin{table*}
\centering
\caption{High-$\tau$ $\rm 9.7\,\mu m$ sources not included in our sample because of a high PAH EW}
\label{rejected}
\begin{tabular}{ccccccccc}
\hline
Name       &  $\alpha$   & $\delta$    & z         & $\tau$ & PAH $\rm EW$   & $\rm log[L_{6}]$ & $\rm log[L_{X}]$    \\
 (1)       &    (2)      & (3)         & (4)       &   (5)  & (6)            & (7)                     & (8)                        \\
\hline
GN\_IRS-4  & 12:36:53.4 & +62:11:39	&  1.27$^2$  &    1.12 &   0.67  &  11.40  & 41.8              \\
GN\_IRS-11 & 12:36:21.3  & +62:17:08     & 1.992$^1$ &  1.55   &  1.18 & 11.41   &  $<$42.8    \\
GN\_IRS-13 & 12:37:34.5 & +62:17:23 &  0.64$^2$ & 	 1.37  &  0.48 &  11.27    &   41.3        \\
GN\_IRS-14 & 12:36:22.5 & +62:15:44  &  0.639$^2$ & 3.00 &   0.63 &  10.39    & 	   41.4   \\
GN\_IRS-15 & 12:37:11.4 & +62:13:31  &  1.99$^2$   &  1.09 &    0.78 &  10.74    &     42.8   \\
GN\_IRS-16 & 12:36:37.0  & +62:08:52   & 2.02$^2$    & 3.97   &    0.57&  11.27   & 42.1               &    \\
GN\_IRS-20 & 12:36:03.2 & +62:11:10  &  0.64$^2$  & 2.32 &    0.60 &  10.68     &      41.5  \\
GN\_IRS-21 & 12:36:18.3  & +62:15:50   & 2.00$^2$  & 1.96     &  0.66 & 11.19    &  42.4 \\
FLS-283      & 17:14:58.3   & +59:24:11   &	0.94$^2$  &  1.12 &    0.68 &  11.46    &      $<$43.1    \\
\hline
\end{tabular}
\begin{list}{}{}
\item The columns are: (1) Name
                       (2),(3) Equatorial {\it Spitzer} coordinates
                       (4) Redshift: $^1$ \citet{Chapman2005};
                                     $^2$ IRS from \citet{Pope2008}, \citet{MurphyIR2009}
                                      in the case of GOODS 
                                     or \citet{Sajina2007} in the case of FLS.
                                              (5) optical depth at 9.7$\mu$m.
                       (6)  PAH equivalent-width at 6.2$\rm \mu m$,
                        in units of \micron (except in the case of FLS-283 
                         where the 11.3 \micron EW is quoted). 
                        (7) Logarithm of the $\nu L_\nu$ monochromatic
                           luminosity at $\rm 6\,\mu m$ in units of solar
                           luminosity as derived by IRS spectroscopy.
                       (8) X-ray luminosity (uncorrected for obscuration) in the 2-10 keV band in units \lunits. 
                            \end{list}
\end{table*}

\section{Discussion}

\subsection{The efficiency of the high-$\tau$ method for finding Compton-thick AGN}

 Silicate absorption features in a few (U)LIRGS in the local
Universe known to be associated with Compton-thick AGN
\citep[e.g. NGC\,6240;][]{Armus2006}, prompts us to investigate whether a high
optical depth at $\rm 9.7\,\mu m$ could provide a reliable diagnostic for the
presence of these nuclei. We can reliably identify Compton-thick AGN in the
nearby Universe by means of X-ray spectroscopy, owing to their high X-ray
brightness. We argue that six out of nine highly absorbed sources at
$\rm 9.7\,\mu m$ ($\tau_{9.7}>1$), in the $\rm 12\,\mu m$ {\it IRAS} Seyfert
sample, with X-ray spectra available in the literature are probably
associated with Compton-thick nuclei.  In any case, all nine sources are heavily 
 obscured AGN. 

Since these are nearby AGN ($z<0.06$),  they should contribute only a small fraction 
 to the X-ray background.  The peak to  the X-ray background is produced 
  at $z\sim 0.7-1 $   from objects with intrinsic luminosities around the `knee' 
   of  the luminosity function, i.e., $\sim10^{43}-10^{44}$ \lunits \citep{Gilli2007}. 
The median X-ray luminosity (uncorrected for absorption) of our six local Compton-thick AGN 
 is $\sim 0.5 \times10^{42}$ \lunits. Assuming a correction of about 33 for the 
  intrinsic X-ray luminosity \citep{Comastri2004},  this translates to $\approx 2\times 10^{43}$ \lunits. 
   Hence, AGN with comparable luminosity, but at higher redshift, are the typical 
    AGN responsible for the peak of the X-ray background. 
 
 At higher redshift, the fraction of candidate  Compton-thick AGN is more uncertain. 
  This fraction can be 
  derived indirectly from the X-ray to IR luminosity ratio. 
  If we take only the X-ray detections into account, this 
  is close to  66\%.  When we also consider the upper limits, 
   we find that the uncertainty becomes large: the fraction of Compton-thick sources 
    could vary between 33\% and 83\%. 
 
The frequency of Compton-thick AGN  measured above  is comparable to or higher 
 than derived by 
other mid-IR methods for finding Compton-thick sources. For example, on the
basis of X-ray stacking analyses, several authors
\citep[e.g.][]{Daddi2007,Fiore2008,Georgantopoulos2008} proposed that the 
$\rm 24\,\mu m$-excess sources are good candidates for hosting 
Compton-thick nuclei. However, because of the lack of X-ray spectroscopy
the fraction of Compton-thick sources cannot be readily quantified. 
Based on brighter sources with X-ray spectroscopy in the GOODS
\citep{Georgantopoulos2011}, AEGIS \citep{Georgakakis2010} and SWIRE fields
\citep{Lanzuisi2009}, the fractions of Compton-thick sources found among IR-excess AGN
 range between 0\% and 50\%. Another widely applied mid-IR method
for the selection of Compton-thick AGN \citep{Alexander2008} relies on 
 low X-ray to IR ratios. The principle behind this method is that 
a low $\rm L_X/L_{IR}$ ratio is a sign of heavy obscuration
\citep[but see][]{Yaqoob2010, Georgakakis2010}. 

Although we have shown that the success rate of our technique in finding
Compton-thick AGN is close to 70\%, at least among optically confirmed Seyferts in the
local Universe, there are two caveats. First, the detection of this feature
alone does not imply the presence of an AGN in these systems. There are
star-forming systems with large absorption features at $\rm 9.7\,\mu m$ that
are associated instead with star-forming galaxies. One case is NGC\,3628 for which 
$\tau_{9.8}=1.64$ \citep{Brandl2006}, although the X-ray emission in this
system does not come from an AGN \citep{Dahlem1996}. In other words, the high
$\tau_{9.7}$ optical depth characterises systems with large amounts of dust but
does not necessarily prove the presence of an AGN. The second caveat has to do
with the completeness of our method. There are well-known Compton-thick sources 
in the literature that contain no significant $\rm 9.7\,\mu m$ absorption. 
We discuss this in detail in the following section. 
 
\subsection{Si optical depth versus X-ray column density}

\begin{figure}
\begin{center}
\includegraphics[width=9.cm]{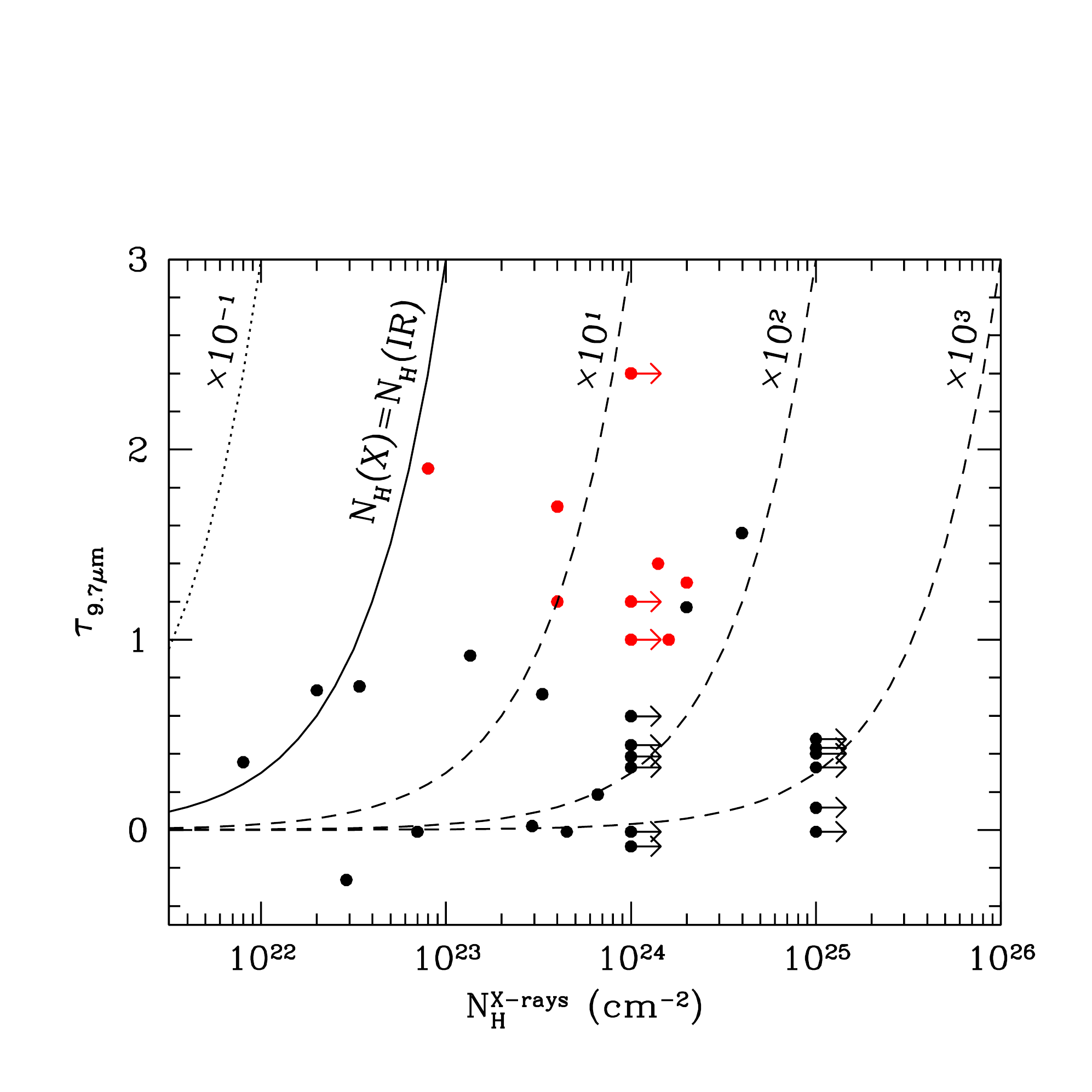} 
\caption{Optical depth at $\rm 9.7\,\mu m$ versus X-ray hydrogen column density
         for the \citet{Shi2006} sample (black points) and the optically-thick
         AGN in the IRAS sample of \citet{Wu2009} (red points). The solid line
         marks the region where the X-ray $N_{\rm H}$ equals that inferred from
         $\tau_{9.7}$, assuming a Milky-Way extinction curve.}
\label{shi}
\end{center}
\end{figure}

\citet{Shi2006} described observations of $\rm 9.7\,\mu m$ features in 97 AGN
including a wide range of types. They found that the strength of the silicate
feature correlates with the hydrogen column density, derived from X-ray
observations in the sense that the low column densities  correspond to silicate emission,
while high columns correspond to silicate absorption. They point out that the
column densities derived from X-ray spectral fitting are always higher than
those estimated on the basis of the Si optical depths. Moreover, there appears
to be a large scatter between the X-ray and the $\rm 9.7\,\mu m$ column density
in the sense that the same $\tau_{9.7}$ corresponds to X-ray column densities
that could differ by as much as an order of magnitude. This result can be
 seen more clearly in Fig.\,\ref{shi}, where we plot the $\rm 9.7\,\mu m$
optical depth as a function of the X-ray column density for both the
$\rm 12\,\mu m $ sample as well as the sample of \citet{Shi2006}. We note that the
latter is based mainly on the optically selected sample of
\citet*{Risaliti1999}. Some column densities in this sample were 
revised by \citet{Malizia2009} and we include these updated values in
Fig.\,\ref{shi}. This figure clearly shows that the majority of the optically
thick sources detected in the mid-IR display Compton-thick absorption at X-ray
wavelengths. In addition we see that there are numerous Compton-thick AGN
according to X-ray spectroscopy that have optical depths $\tau_{9.7}<1$. 

Perhaps the most well studied case of a low Si $\tau$ Compton-thick source is
NGC\,1068. The X-ray column density of this Seyfert-2 galaxy is
$\rm N_H>10^{25}$\,\cunits as derived from high energy X-ray spectroscopy
\citep{Matt2004} with the \sax mission. However, the mid-IR spectrum shows no
prominent $\rm 9.7\,\mu m$ absorption \citep{Sturm2000}. In \citet{Shi2006}
there are many more examples of Compton-thick nuclei which are optically thin at
$\rm 9.7\,\mu m$ (see Fig.\,\ref{shi}). From the above discussion, it becomes
clear that although the presence of a high$-\tau$ absorption feature implies
the presence of a Compton-thick AGN, with a probability of roughly 70\%, the
opposite is not true, i.e., there are many Compton-thick AGN which are not
significantly absorbed at $\rm 9.7\,\mu m$.

The reason for this could be that the X-ray and infrared emitting regions are
not physically coincident and the infrared arises from a region further away from
the accretion disk. It is possible that within the framework
of a clumpy torus model \citep*[see e.g.][]{Nenkova2008,Wada2009,Nikutta2009},
the X-ray emission from the AGN could be obscured by a circumnuclear cloud,
while the infrared emission, which is produced at larger distances from the AGN, 
could experience low or no obscuration at all
\citep[see Fig.\,5 in][]{Shi2006}. 
  \citet{Spoon2007} proposed that the depth of the 
 Si 9.7$\rm \mu m$ feature can be used as a tool to 
  predict the obscuring dust distribution. 
 In this scenario, high silicate depths correspond to 
 sources embedded in a smooth dust distribution, 
  while  low depths correspond to those covered  
 by patchy absorbers.

\subsubsection{Comparison with previous results}
 We to compare our results with the work of \citet{Alexander2008}, 
who searched for Compton-thick AGN in the CDF-N using both IRS spectroscopy in
combination with the $\rm L_X/L_{6 \mu m}$ ratio. \citet{Alexander2008} discussed 
the properties of two sources with available mid-IR IRS spectroscopy in the
CDF-N, claiming that both could be associated with Compton-thick AGN. One
(HDF-oMD49) is an optically identified AGN that is formally undetected in the
2\,Ms data \citet{Alexander2003}. The {\it Spitzer} IRS spectrum and spectral
energy distribution of this object is AGN dominated. On the basis of the source's low
$\rm L_X/L_{IR}$ ratio, \citet{Alexander2008} argue that this AGN is
Compton-thick. The other source (SMMJ123600+621047) is a z=2.002
submillimeter-emitting galaxy (SMG) with a mid-IR-bright AGN \citep{Pope2008}
that is again undetected in the 2\,Ms CDF-N data. The $\rm L_X/L_{6 \mu m}$
luminosity ratio indicates that it hosts a Compton-thick AGN, although optical
and near-IR spectroscopic observations do not reveal the signatures of AGN
activity, leaving open the possibility that this source is associated with a
star-forming galaxy and not with an AGN. None of these sources display a
$\tau>1$ 9.7$\mu m$ absorption feature in its IRS spectrum.

\section{Summary}
The goal of this work has been to investigate whether the presence of high-absorption 
at mid-IR wavelengths ($\tau_{9.7} >1 $) can be used as an efficient 
 means for identifying highly obscured or even Compton-thick 
AGN. Our conclusions can be summarised as follows:
\begin{itemize}
\item{Our re-analysis of {\it Spitzer}-IRS spectra shows that there are eleven AGN 
      with $\tau_{9.7\mu m}>1$ in the 12$\mu m$ Seyfert {\it IRAS} sample. For
      nine sources X-ray spectroscopic observations are available in the
      literature. The X-ray spectra show that a large fraction (six of nine) is
      probably associated with Compton-thick AGN. The success rate of this method is
      comparable to or even higher than those of other methods using either mid-IR
      photometric information or the X-ray to IR luminosity ratio.}
\item{Even though optical thickness in the MIR, as determined by 9.7 micron Si
      absorption with $\tau>1$ can be efficiently used to identify
      Compton-thick sources in the local Universe with a high success rate,
      this method cannot provide complete samples of these sources. This is
      because a large fraction of Compton-thick sources do not have large high Si
      optical depths. This could possibly be explained within the framework of a
      clumpy torus model.}
\item{Using {\it Spitzer}-IRS spectra in the CDF and the FLS surveys, we
      have compiled a sample of 12 sources in total, with $\tau_{9.7} >1$ at higher
      redshift (z=0.87-2.70). These are selected to have weak PAH features 
        to maximise the probability that they host an AGN. The vast majority of 
       sources in the CDFs (six out of seven) are detected in the X-rays, while all five FLS sources
      remain undetected. For the X-ray detections, the high
      X-ray luminosities ($\rm >2\times10^{42}$\,\lunits) confirm that all these  
      sources are AGN.  For two FLS sources, their optical line
      emission also indicates that an AGN is present.  
      Because our sources are faint in X-rays, we are unable to ascertain 
        from X-ray spectroscopy whether 
         at least some of our sources are Compton-thick.   
      The use of the $\rm L_X/L_{6\mu m}$ ratio
      indicates that the success  rate of the high-$\tau$ method 
       in identifying  Compton-thick AGN 
       could be  about 66 \% (4/6), if we consider X-ray detections only.
        However, it can be as low as $\sim$33 \% (4/12), or as high as 
        83\% under the extreme 
         assumptions that all the X-ray upper limits are either  associated 
          or not associated with Compton-thick sources. 
    Deeper X-ray spectroscopy will be needed to
      properly constrain the exact fraction of Compton-thick AGN among high-z,
      high-$\tau$ sources.}
      \end{itemize}
      In the future, observations with the {\it JWST} 
      will be able to identify numerous galaxies with deep Si features that 
      could be used to compile large candidate Compton-thick AGN samples.

\begin{acknowledgements}
IG and AC acknowledge support by the European Community through the Marie Curie fellowship
FP7-PEOPLE-IEF-2008 Prop. 235285 under the Seventh Framework Programme (FP7/2007-2013). 
 KD also acknowledges support by the European Community through a Marie Curie
Fellowship (PIEF-GA-2009-235038) awarded under the Seventh Framework Programme
(FP7/2007-2013)
The {\it Chandra} data were taken from the {\it Chandra} Data Archive 
at the {\it Chandra} X-ray Center. This work is based on observations made with the {\it Spitzer} 
 Space Telescope, which is operated by the Jet Propulsion Laboratory, 
 California Institute of Technology under contract with NASA.
\end{acknowledgements}

\end{document}